\documentclass[fleqn,usenatbib]{mnras}

\usepackage{newtxtext,newtxmath}

\usepackage[T1]{fontenc}

\DeclareRobustCommand{\VAN}[3]{#2}
\let\VANthebibliography\thebibliography
\def\thebibliography{\DeclareRobustCommand{\VAN}[3]{##3}\VANthebibliography}

\usepackage{graphicx}	% Including figure files
\usepackage{amsmath}	% Advanced maths commands
\usepackage{amssymb}	% Extra maths symbols
\usepackage{booktabs}
\usepackage{multirow}

\title[Imprints of the first stars on the NIRB]{Revealing the formation histories of the first stars with the cosmic near-infrared background}

\author[G.~Sun et al.]{Guochao~Sun$^{1}$\thanks{E-mail: gsun@astro.caltech.edu}, Jordan~Mirocha$^{2}$, Richard~H.~Mebane$^{3,4}$, Steven~R.~Furlanetto$^{4}$ \\
$^{1}$Cahill Center for Astronomy and Astrophysics, California Institute of Technology, 1200 E California Blvd, Pasadena, CA 91125, USA \\
$^{2}$McGill University, Department of Physics \& McGill Space Institute, 3600 Rue University, Montr\'eal, QC, H3A 2T8 \\
$^{3}$Department of Astronomy and Astrophysics, University of California, Santa Cruz, 1156 High Street, Santa Cruz, CA 95064, USA \\
$^{4}$Department of Physics and Astronomy, University of California, Los Angeles, CA 90024, USA}

\date{Accepted XXX. Received YYY; in original form ZZZ}

\pubyear{2021}

\begin{document}

\defcitealias{FZ_2013}{FZ13}

\label{firstpage}
\pagerange{\pageref{firstpage}--\pageref{lastpage}}
\maketitle

\begin{abstract}
The cosmic near-infrared background (NIRB) offers a powerful integral probe of radiative processes at different cosmic epochs, including the pre-reionization era when metal-free, Population III (Pop III) stars first formed. While the radiation from metal-enriched, Population II (Pop II) stars likely dominates the contribution to the observed NIRB from the reionization era, Pop III stars --- if formed efficiently --- might leave characteristic imprints on the NIRB thanks to their strong Ly$\alpha$ emission. Using a physically-motivated model of first star formation, we provide an analysis of the NIRB mean spectrum and anisotropy contributed by stellar populations at $z>5$. We find that in circumstances where massive Pop III stars persistently form in molecular cooling haloes at a rate of a few times $10^{-3}\,M_\odot \ \mathrm{yr}^{-1}$, before being suppressed towards the epoch of reionization (EoR) by the accumulated Lyman-Werner background, a unique spectral signature shows up redward of $1\,\mu$m in the observed NIRB spectrum sourced by galaxies at $z>5$. While the detailed shape and amplitude of the spectral signature depend on various factors including the star formation histories, IMF, LyC escape fraction and so forth, the most interesting scenarios with efficient Pop III star formation are within the reach of forthcoming facilities such as the Spectro-Photometer for the History of the Universe, Epoch of Reionization and Ices Explorer (SPHEREx). As a result, new constraints on the abundance and formation history of Pop III stars at high redshifts will be available through precise measurements of the NIRB in the next few years.
\end{abstract}


%%%%%%%%%%%%%%%%%%%%%%%%%%%%%%%%%%%%%%%%%%%%%%%%%%

%%%%%%%%%%%%%%%%% BODY OF PAPER %%%%%%%%%%%%%%%%%%

% -------------------------- S1: Introduction -------------------------- %

\section{Introduction}

Population~III (Pop~III) stars are believed to form in primordial, metal-free gas clouds cooled via molecular hydrogen ($\mathrm{H_2}$) at very high redshift, well before metal-poor, Population~II (Pop~II) stars typical for distant galaxies started to form. These first generation of stars at the so-called cosmic dawn were responsible for the onset of cosmic metal enrichment and reionization, and their supernova remnants may be the birthplaces of supermassive black holes observed today \cite[see recent reviews by][]{Bromm_2013, IVH_2020}. Despite their importance in understanding the cosmic history of star formation, Pop~III stars are incredibly difficult to directly detect, even for the upcoming generation of telescopes like the \textit{James Webb Space Telescope} (JWST) as discussed in \citet{Rydberg_2013} and \citet{Schauer_2020ApJ}, and thus constraints on their properties remain elusive. Nevertheless, the formation and physical properties of Pop~III stars have been investigated in detail with theoretical models over the past few decades, and several promising observing methods have been proposed to discover them in the near future. 

Theoretical models of Pop~III stars come in many forms, including simple analytical arguments \cite[e.g.,][]{MT_2008}, detailed numerical simulations \cite[e.g.,][]{ABN_2002, WA_2007, ON_2007,  Maio_2010, Greif_2011, Safranek-Shrader_2012, SGB_2012, Xu_2016}, and semi-analytic models that balance computational efficiency and physical accuracy \cite[e.g.,][]{TS_2009, TSS_2009, Crosby_2013, Jaacks_2018, Mebane_2018, VHB_2018, LB_2020} These theoretical efforts reveal a detailed, though still incomplete, picture of how the transition from Pop~III to metal-enriched, Pop~II star formation might have occurred. Minihaloes above the Jeans/filtering mass scale set by some critical fraction of $\mathrm{H_2}$ \citep{Tegmark_1997} and below the limit of atomic hydrogen cooling are thought to host the majority of Pop~III star formation since $z\gtrsim30$, where the rotational and vibrational transitions of collisionally-excited $\mathrm{H_2}$ dominate the cooling of primordial gas\footnote{Stars formed out of primordial gas in these molecular cooled haloes are sometimes referred to as Pop~III.1 stars, whereas stars formed in atomic cooling haloes that are primordial but affected by previously-generated stellar radiation are referred to as Pop~III.2 stars.}. The lack of efficient cooling channels yields a Jeans mass of the star-forming region as high as a few hundred $M_\odot$, producing very massive and isolated Pop~III stars in the classical picture \citep{BL_2004}. However, simulations indicate that even modest initial angular momentum of the gas in minihaloes could lead to fragmentation of the protostellar core and form Pop~III binaries or even multiple systems \cite[e.g.,][]{TAO_2009, SGB_2010, Sugimura_2020}, which further complicates the Pop~III initial mass function (IMF). Several physical processes contribute to the transition to Pop~II star formation.  The feedback effect of the Lyman-Werner (LW) radiation background built up by the stars formed is arguably consequential for the formation of Pop~III stars. LW photons ($11.2\,\mathrm{eV}<h\nu<13.6\,\mathrm{eV}$) can regulate Pop~III star formation by photo-dissociating $\mathrm{H_2}$ through the two-step Solomon process \citep{SW_1967} and thereby setting the minimum mass of minihaloes above which Pop~III stars can form \cite[][]{HRL_1997, WGHB_2011, HF_2012, SGB_2012, Visbal_2014, Mebane_2018}, although some recent studies suggest that $\mathrm{H_2}$ self-shielding might greatly alleviate the impact of the LW background \cite[see e.g.,][]{SW_2020}. Other important factors to be considered in modelling the transition include the efficiency of metal enrichment (i.e., chemical feedback) from Pop~III supernovae \cite[][]{Pallottini_2014, SSC_2018}, the X-ray background sourced by Pop~III binaries that might replenish $\mathrm{H_2}$ by catalyzing its formation \cite[][]{HAR_2000, Hummel_2015, Ricotti_2016}, and the residual streaming velocity between dark matter and gas \cite[][]{TH_2010, NYG_2012, Fialkov_2012, Schauer_2020arXiv}. In spite of all the theoretical efforts, substantial uncertainties remain in how long and to what extent Pop~III stars might have coexisted with their metal-enriched descendants, leaving the timing and duration of the Pop~III to Pop~II transition largely unconstrained. 

Direct constraints on Pop~III stars would be made possible by detecting their emission features. One such feature is the $\ion{He}{II}$ $\lambda$1640 line, which is a strong, narrow emission line indicative of a very hard ionizing spectrum typical for Pop~III stars \cite[][]{Schaerer_2003}. The association of the $\ion{He}{II}$ $\lambda1640$ line with Pop~III stars has been pursued in the context of both targeted observations \cite[e.g.,][]{Nagao_2005, Cai_2011, MDF_2016} and statistical measurements via the line-intensity mapping technique \cite[e.g.,][]{VHB_2015}. While possible identifications have been made for objects such as ``CR7'' \cite[][]{Sobral_2015}, the measurements are controversial and a solid $\ion{He}{II}$ $\lambda1640$ detection of Pop~III stars may not be possible until the operation of next-generation ground-based telescopes such as the E-ELT \cite[][]{Grisdale_2021}. A number of alternative (and often complementary) probes of Pop~III stars have therefore been proposed, including long gamma-ray bursts (GRBs) associated with the explosive death of massive Pop~III stars \cite[][]{MR_2010,TSM_2011}, caustic transits behind lensing clusters \cite[][]{Windhorst_2018}, the cosmic near-infrared background \cite[NIRB,][]{SBK_2002, Kashlinsky_2004,FZ_2013,Yang_2015,Helgason_2016,Kashlinsky_2018RvMP}, and spectral signatures in the global 21-cm signal \cite[][]{TZ_2008, Fialkov_2014, Mirocha_2018, Mebane_2020} and 21-cm power spectrum \citep{Fialkov_2013, Fialkov_2014, Qin_2021}. 

Pop~III stars have been proposed as a potential explanation for the observed excess in the NIRB fluctuations \cite[][]{SF_2003, Kashlinsky_2004, Kashlinsky_2005}, which cannot be explained by the known galaxy populations with sensible faint-end extrapolation \citep{Helgason_2012}, and their accreting remnants provide a viable explanation for the coherence between the NIRB and the soft cosmic X-ray background (CXB) detected at high significance \cite[][]{Cappelluti_2013}. However, subsequent studies indicate that, for Pop~III stars to source a considerable fraction of the observed NIRB, their formation and ionizing efficiencies would need to be so extreme that constraints on reionization and the X-ray background are likely violated \cite[e.g.,][]{MS_2005, Helgason_2016}. Consequently, some alternative explanations have been proposed, such as the intrahalo light (IHL) radiated by stars stripped away from parent galaxies during mergers \cite[][]{Cooray_2012Nat, Zemcov_2014Sci}, with a major contribution from sources at $z<2$, and accreting direct collapsed black holes (DCBHs) that could emit a significant amount of rest-frame, optical--UV emission at $z\gtrsim12$ due to the absorption of ionizing radiation by the massive accreting envelope surrounding them \cite[][]{Yue_2013_DCBH}. 

Pop~III stars alone are likely insufficient to fully explain the source-subtracted NIRB fluctuations observed and separating their contribution to the NIRB from other sources, including Pop~II stars that likely co-existed with Pop~III stars over a long period of time, will be challenging. Nevertheless, there is continued interest in understanding and modelling potential signatures of Pop~III stars in the NIRB \cite[e.g.,][]{Kashlinsky_2004, Kashlinsky_2005, Yang_2015, Helgason_2016}, which is one of only a few promising probes of Pop~III in the near term. In particular, Fernandez and Zaroubi (\citeyear{FZ_2013}, hereafter \citetalias{FZ_2013}) point out that strong Ly$\alpha$ emission from Pop~III stars can lead to a ``bump'' in the mean spectrum of the NIRB, a spectral signature that can reveal information about physical properties of Pop~III stars and the timing of the Pop~III to Pop~II transition. The soon-to-be-launched satellite Spectro-Photometer for the History of the Universe, Epoch of Reionization and Ices Explorer (SPHEREx; \citealt{Dore_2014}) has the raw sensitivity to detect the contribution of galaxies during the epoch of reionization (EoR) to the NIRB at high significance \citep{Feng_2019}, making it possible, at least in principle, to detect or rule out such spectral features. However, despite significant differences in detailed predictions, previous modelling efforts \cite[e.g.,][]{FK_2006, Cooray_2012, Yue_2013, Helgason_2016} have suggested that first galaxies during and before the EoR may only contribute to approximately less than 1\% of both the source-subtracted NIRB mean intensity and its angular fluctuations, as measured from a series of deep imaging surveys \cite[e.g.,][]{Kashlinsky_2012, Zemcov_2014Sci, Seo_2015}. A challenging measurement notwithstanding, unprecedented NIRB sensitivities of space missions like SPHEREx and the Cosmic Dawn Intensity Mapper (CDIM; \citealt{Cooray_2019BAAS}) urge the need for an improved modelling framework to learn about the first galaxies from future NIRB measurements. 

In this work, we establish a suite of NIRB predictions that are anchored to the latest constraints on the high-$z$ galaxy population drawn from many successful \textit{Hubble Space Telescope} (HST) programs, such as the Hubble Ultra Deep Field \citep{Beckwith_2006HUDF}, CANDELS \citep{Grogin_2011CANDELS}, and Hubble Frontier Fields \citep{Lotz_2017FF}. We employ a semi-empirical model to describe the known galaxy population, and then add in a physically-motivated, but flexible, model for Pop~III stars that allow us to explore a wide range of plausible scenarios. This, in various aspects, improves over previous models, which, e.g., parameterized the fraction of cosmic star formation in Pop~III haloes as a function of redshift only and/or employed simpler Pop~II models calibrated to earlier datasets (e.g., \citealt{Cooray_2012}; \citetalias{FZ_2013}; \citealt{Helgason_2016}; \citealt{Feng_2019}). These advancements not only allow more accurate modelling of the contribution to the NIRB from high-$z$ galaxies, but also provide a convenient physical framework to analyse and interpret datasets of forthcoming NIRB surveys aiming to quantify the signal level of galaxies during and before reionization. 

This paper is organized as follows. In Section~\ref{sec:modelling}, we describe how we model the spatial and spectral properties of the NIRB associated with high-$z$ galaxies, using a simple, analytical framework of Pop~II and Pop~III star formation in galaxies at $z>5$. We present our main results in Section~\ref{sec:results}, including the predicted NIRB signals, potential spectral imprints due to Pop~III star formation, and sensitivity estimates for detecting Pop~II and Pop~III signals in future NIRB surveys. In Section~\ref{sec:imply}, we show implications for other observables of high-$z$ galaxies that can be potentially drawn from NIRB observations. We discuss a few important caveats and limitations of our results in Section~\ref{sec:discussion}, before briefly concluding in Section~\ref{sec:conclusions}. Throughout this paper, we assume a flat, $\Lambda$CDM cosmology consistent with the results from the \cite{Planck2016XIII}. 

\begin{table*}
\centering
\caption{Parameter values in the reference models of Pop~II and Pop~III star formation.}
\begin{tabular}{cccccccccc}
\hline
Symbol & Parameter & Reference & Model I & Model II & Model III & Model A & Model B & Model C & Model D \\
\hline
\multicolumn{9}{c}{\textit{Pop~II stars}} \\
\cmidrule(l{20em}r{20em}){1-8}
$f_*$ & star formation efficiency & equation~(\ref{eq:sfe-dpl}) & dpl & steep & floor \\
$Z$ & stellar metallicity & Section~(\ref{sec:stellar_cont}) & 0.02 & 0.02 & 0.02 \\
$f^{\rm II}_{\rm esc}$ & LyC escape fraction & equation~(\ref{eq:Jnu0}) & 0.1 & 0.1 & 0.1 \\
$f^{\rm II}_{\rm esc, LW}$ & LW escape fraction & Section~(\ref{sec:sfh_pop3}) & 1 & 1 & 1\\
\multicolumn{9}{c}{\textit{Pop~III stars}} \\
\cmidrule(l{20em}r{20em}){1-8}
$Q(\mathrm{H})$ [$\mathrm{s}^{-1}$] & H photoionization rate & equation~(\ref{eq:QH}) & & & & $10^{50}$ & $10^{51}$ & $10^{50}$ & $10^{51}$ \\
$\dot{M}^{\rm III}_{*}$ [$M_{\odot}\,\mathrm{yr}^{-1}$] & SFR per halo & equation~(\ref{eq:sfrd_pop3}) & & & & $1\times10^{-3}$ & $2\times10^{-4}$ & $3\times10^{-6}$ & $1\times10^{-5}$ \\
$\mathcal{T}_c$ [$\mathrm{Myr}$] & critical time limit & Section~(\ref{sec:sfh_pop3}) & & & & 25 & 0 & 0 & 250 \\
$\mathcal{E}_c$ [$\mathrm{erg}$] & critical binding energy & Section~(\ref{sec:sfh_pop3}) & & & & $3\times10^{52}$ & $8\times10^{51}$ & $1\times10^{52}$ & $5\times10^{52}$ \\
$f^{\rm III}_{\rm esc}$ & LyC escape fraction & Fig.~(\ref{fig:fprofile}) & & & & 0.05/0.2 & 0.05/0.2 & 0.05/0.2 & 0.05/0.2 \\
$f^{\rm III}_{\rm esc, LW}$ & LW escape fraction & Section~(\ref{sec:sfh_pop3}) & & & & 1 & 1 & 0 & 1 \\
\hline
\end{tabular}
\label{tb:model_params}
\end{table*}

% -------------------------- S2: Models -------------------------- %

\section[]{Models} \label{sec:modelling}

\subsection{Star formation history of high-redshift galaxies} \label{sec:sfhs}

\subsubsection{The formation of Pop~II stars}
\label{sec:sfh_pop2}

Following \citet{Mirocha_2017}, we model the star formation rate density (SFRD) of normal, high-$z$ galaxies as an integral of the star formation rate (SFR) per halo $\dot{M}_*(M_h)$ over the halo mass function $n(M_h)$ \cite[see also][]{SF_2016, Furlanetto_2017}
\begin{align}
\dot{\rho}^\mathrm{II}_{*}(z) & = \int_{M^\mathrm{II}_{h,\mathrm{min}}} n(M_h) \dot{M}_{*}(M_h, z) d M_h \nonumber \\
& = \int_{M^\mathrm{II}_{h,\mathrm{min}}} n(M_h) f_*(M_h, z) \frac{\Omega_b}{\Omega_m} \dot{M}_h(M_h, z) d M_h~,
\end{align}
where $M^\mathrm{II}_{h,\mathrm{min}}$ is generally evaluated at a virial temperature of $T_{\rm vir}=10^4\,$K, a free parameter in our model above which Pop~II are expected to form due to efficient cooling via neutral atomic lines \cite[][]{OH_2002}, namely $M^\mathrm{II}_{h,\mathrm{min}}=M^\mathrm{III}_{h,\mathrm{max}}$. $\dot{M}_*(M_h)$ is further specified by a star formation efficiency (SFE), $f_*$, defined to be the fraction of accreted baryons that eventually turn into stars, and the mass growth rate, $\dot{M}_h$, of the dark matter halo. We exploit the abundance matching technique to determine the mean halo growth histories by matching halo mass functions at different redshifts. As illustrated in \citet{Furlanetto_2017} and \citet{MLPL_2020}, the abundance-matched accretion rates given by this approach are generally in good consistency with results based on numerical simulations \citep{TCM_2015} for atomic cooling haloes at $5 \lesssim z \lesssim 10$ (but see \citealt{SGM_2021} for a comparison with estimates based on the extended Press-Schechter formalism). Even though effects like mergers and the stochasticity in $\dot{M}_h$ introduce systematic biases between the inferences made based on merger trees and abundance matching, such biases can be largely eliminated by properly normalizing the nuisance parameters in the model \citep{MLPL_2020}. By calibrating to the latest observational constraints on the galaxy UV luminosity function (UVLF), \citet{Mirocha_2017} estimate $f_*$ to follow a double power-law in halo mass (the \texttt{dpl} model)
\begin{equation}
f^{\rm dpl}_*(M_h) = \frac{f_{*,0}}{\left( \frac{M_h}{M_{\rm p}} \right)^{\gamma_{\rm lo}} + \left( \frac{M_h}{M_{\rm p}} \right)^{\gamma_{\rm hi}}}~,
\label{eq:sfe-dpl}
\end{equation}
with no evident redshift evolution, in agreement with other recent work \citep[e.g.,][]{MTT_2015, Tacchella_2018, Behroozi_2019, Stefanon_2021arXiv}. The evolution of $f_*$ for low-mass haloes is however poorly constrained by the faint-end slope of the UVLF, and can be highly dependent on the regulation of feedback processes \citep{Furlanetto_2017, Furlanetto_2021} and the burstiness of star formation \citep{FM_2021arXiv}. Therefore, in addition to the baseline \texttt{dpl} model, we consider two alternative parameterization --- one suggested by \citet{Okamoto2008} that allows a steep drop of $f_*$ for low-mass haloes (the \texttt{steep} model)
\begin{equation}
f^{\rm steep}_*(M_h) = \left[ 1 + \left(2^{\mu/3} - 1\right) \left(\frac{M_h}{M_{\rm crit}}\right)^{-\mu} \right]^{-3/\mu}~,
\label{eq:sfe-steep}
\end{equation}
and the other that imposes a constant floor on the SFE of 0.005 (the \texttt{floor} model). In this work, we take the same best-fit parameters as those given by \citet{Mirocha_2017} to define the two reference Pop~II models, namely $f_{*,0}=0.05$, $M_{\rm p}=2.8\times10^{11}$, $\gamma_{\rm lo}=0.49$, $\gamma_{\rm hi}=-0.61$, with $\mu=1$ and $M_{\rm crit}=10^{10}\,M_{\odot}$ for the \texttt{steep} model\footnote{The SFE parameters taken are fit to the observed UVLFs measured by \citet{Bouwens_2015} at $6<z<8$, which agree reasonably well with the most recent measurements in \cite[e.g.,][]{Bouwens_2021}.}. With the three variants of our Pop~II SFE model, we aim to bracket a reasonable range of possible low mass/faint-end behaviour, and emphasize that future observations by the JWST \cite[e.g.,][]{Furlanetto_2017, Yung_2019} and line-intensity mapping surveys \cite[e.g.,][]{Park_2020, Sun_2021} can place tight constraints on these models. 

\begin{figure}
 \centering
 \includegraphics[width=0.48\textwidth]{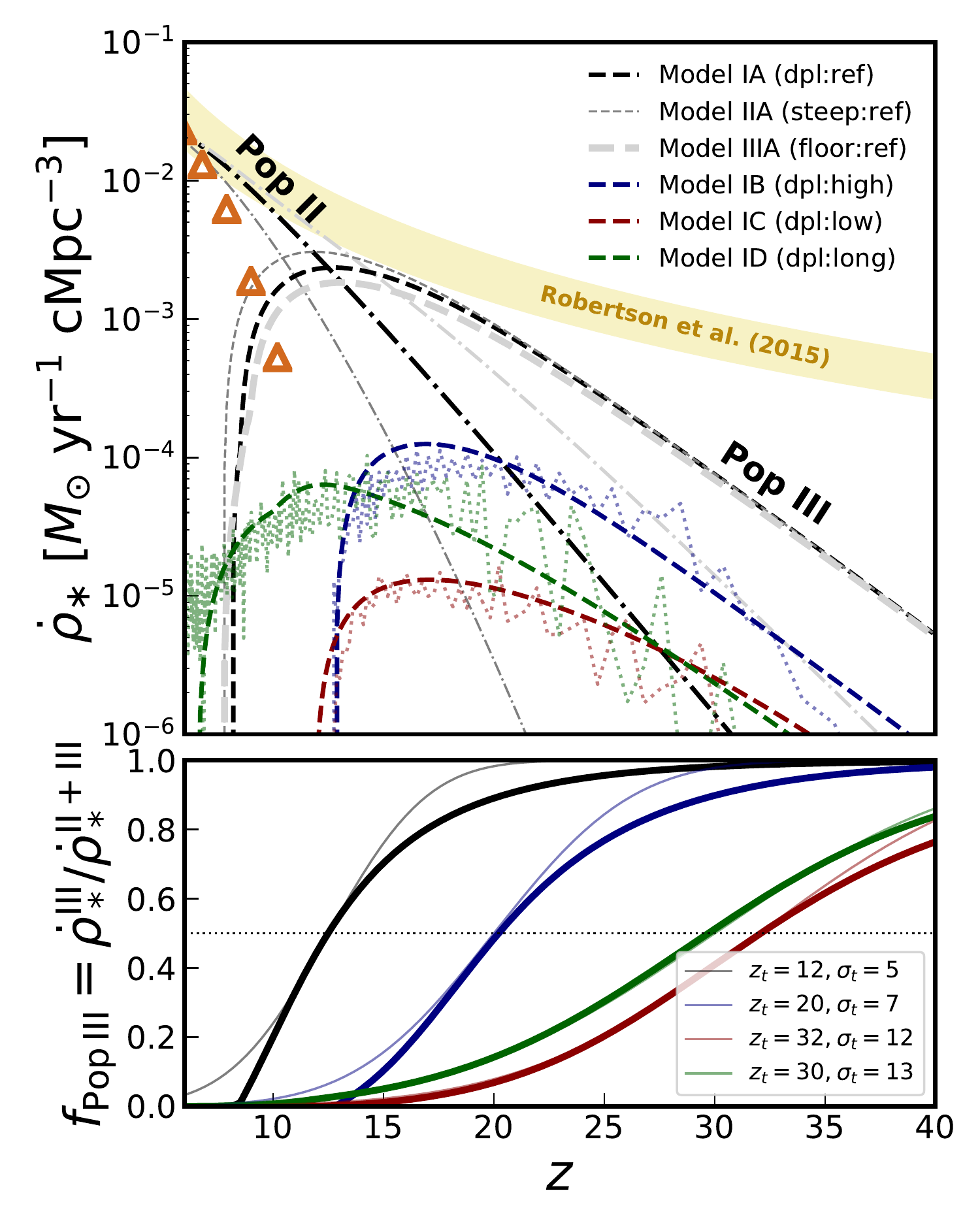}
 \caption{Pop~II and Pop~III star formation histories in different models considered in this work, as specified in Table~\ref{tb:model_params}. \textit{Top:} SFRDs of Pop~II (dash-dotted) and Pop~III (dashed) stars. The black curves represent our reference model (Model~IA), with the thin dark grey curve and the thick light grey curve representing variations where the Pop~II SFE follows the \texttt{steep} (Model~II) and \texttt{floor} (Model~III) models, respectively. The bottom set of three dotted curves show the Pop~III histories derived with the semi-analytical approach in \citet{Mebane_2018}, to which Models~IB, IC, and ID are calibrated. The shaded region and open triangles represent the cosmic SFRD inferred from the maximum-likelihood model by \citet{Robertson_2015} and the observed SFRD (integrated to a limiting SFR of $0.3\,M_\odot\,\mathrm{yr}^{-1}$) up to $z=10$ determined by \citet{Oesch_2018}, respectively. \textit{Bottom:} the stellar population transition represented by the ratio of Pop~III and total SFRDs. For comparison, approximations made with the functional form $f_\mathrm{Pop~III}(z)=1/2+\mathrm{erf}[(z-z_t)/\sigma_t]/2$ are shown by the thin curves.}
 \label{fig:SFRD}
\end{figure}

\subsubsection{The formation of Pop~III stars}
\label{sec:sfh_pop3}

While the star formation history of Pop~II stars may be reasonably inferred by combing existing observational constraints up to $z\sim10$ with physically-motivated extrapolations towards higher redshifts, the history of Pop~III stars is only loosely constrained by observations. Several recent studies \cite[e.g.,][]{Visbal_2014, Jaacks_2018, Mebane_2018, SSC_2018, LB_2020} investigate the formation of Pop~III stars under the influence of a variety of feedback processes, including the LW background and supernovae. In general, these models find that Pop~III SFRD increases steadily for approximately 200\,Myr since the onset of Pop~III star formation at $z\gtrsim30$, before sufficiently strong feedback effects can be established to regulate their formation. In detail, however, the predicted Pop~III SFRDs differ substantially in both shape and amplitude. Massive Pop~III star formation can persist in minihaloes for different amounts of time depending on factors such as the strength of LW background and the efficiency of metal enrichment (which, in turn, depends on how metals can be produced, retained and mixed within minihaloes). Consequently, the formation of Pop~III stars can either terminate as early as $z>10$ in some models, or remain a non-negligible rate greater than $10^{-4}\,\mathrm{M_\odot\,yr^{-1}\,Mpc^{-3}}$ through the post-reionization era in others. Given the large uncertainty associated with the Pop~III SFRD, we follow \citet{Mirocha_2018} and account for the Pop~III to Pop~II transition with a simple descriptive model, which offers a flexible way to simultaneously capture the physics of Pop~III star formation and encompass a wide range of possible scenarios. We defer the interested readers to that paper and only provide a brief summary here. 

We assume that Pop~III stars can only form in minihaloes with halo mass between $M^\mathrm{III}_{h,\mathrm{min}}$ and $M^\mathrm{III}_{h,\mathrm{max}}$ at a constant rate $\dot{M}^\mathrm{III}_*$ per halo, in which case the Pop~III SFRD can be written as
\begin{equation}
\dot{\rho}^\mathrm{III}_*(z) = \dot{M}^\mathrm{III}_* \int^{M^\mathrm{III}_{h,\mathrm{max}}}_{M^\mathrm{III}_{h,\mathrm{min}}} n(M_h) d M_h~.
\label{eq:sfrd_pop3}
\end{equation}
The minimum mass, $M^\mathrm{III}_{h,\mathrm{min}}$, of Pop~III star-forming haloes is set by the threshold for effective $\mathrm{H_2}$ cooling, regulated in response to the growing LW background following \citet{Visbal_2014}. The maximum mass, $M^\mathrm{III}_{h,\mathrm{max}}$, of Pop~III star-forming haloes is controlled by two free parameters, which set the critical amount of time individual haloes spend in the Pop~III phase, $\mathcal{T}_c$, as well as a critical binding energy, $\mathcal{E}_c$, at which point haloes are assumed to transition from Pop~III to Pop~II star formation. The first condition effectively results in a fixed amount of stars (and metals) produced per halo in our model, and thus serves as a limiting case in which the Pop~III to Pop~II transition is governed by the \textit{production} of metals. The second condition enforced by $\mathcal{E}_c$ provides a contrasting limiting case, in which the transition from Pop~III to Pop~II is instead governed by metal \textit{retention}. In practice, $\mathcal{E}_c$ may range from as small as the typical energy output of a supernova ($\sim10^{51}\,$erg) to a few hundred times larger\footnote{As discussed in \citet{Mirocha_2018}, it is likely that $\dot{M}^\mathrm{III}_*$, $\mathcal{T}_c$, and $\mathcal{E}_c$ are actually positively-correlated with each other in reality, but we ignore such subtleties here to maximally explore the possible scenarios.}. It is worth noting that, rather than quantifying the impact of metal enrichment on Pop~III star formation and the corresponding NIRB signal through a global volume-filling factor of metal-enriched IGM due to galactic outflows \cite[see e.g.,][]{Yang_2015}, we use $\mathcal{T}_c$, and $\mathcal{E}_c$ to control the Pop~III to Pop~II transition. Although this approach does not invoke the metallicity of halos explicitly, it is flexible enough to produce SFRDs that are in good agreement with more sophisticated models, which do link the Pop~III to Pop~II transition to halo metallicity \citep[e.g.,][]{Mebane_2018}. Finally, for simplicity, we assume blackbody spectrum for Pop~III stars and scale the ionizing flux with the parameter $Q(\mathrm{H})$, which we describe in more detail in \S\ref{sec:model_mean_la}.

Fig.~\ref{fig:SFRD} shows the star formation histories of Pop~II and Pop~III stars calculated from a collection of models we consider in this work. Values of key model parameters adopted are summarized in Table~\ref{tb:model_params}. Specifically, three different cases  \cite[all permitted by current observational constraints, see e.g.,][]{Mirocha_2017} of extrapolating Pop~II star formation down to low-mass, atomic-cooling haloes unconstrained by the observed UVLFs are referred to as Model~I (\texttt{dpl}, see equation~\ref{eq:sfe-dpl}), Model~II (\texttt{steep}, see equation~\ref{eq:sfe-steep}), and Model~III (\texttt{floor}), respectively. $f_{\rm esc}$ and $f_{\rm esc,LW}$ represent the escape fractions of Lyman continuum (LyC) and LW photons, respectively. Four Pop~III models with distinct SFRDs resulting from different combinations of $\dot{M}^\mathrm{III}_*$, $\mathcal{T}_c$, and $\mathcal{E}_c$ are considered. Model~A represents an optimistic case with extremely efficient formation of massive, Pop~III stars that leads to a prominent signature on the NIRB. To form $100\,M_\odot$ Pop~III stars that yields $Q(\rm H)\sim10^{50}\,\mathrm{s}^{-1}$ at a rate as high as $\dot{M}^\mathrm{III}_* \sim 10^{-3}\,M_\odot\,\mathrm{yr}^{-1}$ in minihaloes with a typical baryonic mass accretion rate of $10^{-3}$--$10^{-2}\,M_\odot\,\mathrm{yr}^{-1}$\cite[e.g.,][]{Greif_2011,Susa_2014}, the star formation efficiency must be exceedingly high and even close to unity over long timescales. This, in turn, requires a relatively inefficient coupling between the growth of Pop~III stars and the radiative and mechanical feedback. Models~B, C, and D are our model approximations to Pop~III histories derived with the semi-analytical approach described in \citet{Mebane_2018}. Similar to Model~A, all these models yield Pop~III SFRDs regulated by LW feedback associated with Pop~II and/or Pop~III stars themselves, as controlled by the parameters $f^{\rm II}_{\rm esc,LW}$ and $f^{\rm III}_{\rm esc,LW}$. We note that setting $f^{\rm III}_{\rm esc,LW}$ to zero (as in Model~C) is only meant to turn the LW feedback off, since in reality the escape fraction of LW photons tends to be order of unity in the far-field limit \cite[see e.g.,][]{Schauer_2017}. Besides the LW feedback that sets the end of the Pop~III era, the amplitude of the Pop~III SFRD is also determined by the prescription of Pop~III star formation. Among the three models, Model~C approximates the scenario where Pop~III stars with a normal IMF form at a low level of stellar mass produced per burst, which yields NIRB signals likely inaccessible to upcoming observations, whereas Models B and D approximate scenarios where Pop~III stars form more efficiently and persistently, respectively, and if massive enough ($M_* \sim 500\,M_\odot$), can leave discernible imprints on the NIRB. For comparison, two additional cosmic SFRDs are shown: (i) that inferred from \citet{Robertson_2015} by integrating the UVLFs down to $L_{\rm UV}\sim0.001\,L_*$ (yellow band), and (ii) that reported in \citet{Oesch_2018} which includes observed galaxies with $\dot{M}_* \gtrsim 0.3\,M_{\odot}\mathrm{yr}^{-1}$ (open triangles). 

To put things into the context of the literature, we show in the lower panel of Fig.~\ref{fig:SFRD} the fraction of stars that are Pop~III at each redshift. Predictions from our models are shown together with approximations made using the functional form
$f_\mathrm{Pop~III}(z)=1/2+\mathrm{erf}[(z-z_t)/\sigma_t]/2$, which is frequently adopted in the literature to estimate the Pop~III contribution \cite[e.g.,][]{Cooray_2012, FZ_2013, Feng_2019}. It can be seen that, compared with the phenomenological description using the error function, our physical models imply a more extended early phase with the Pop~II SFRD gradually catching up. The late-time behaviour is characterized by how sharply the Pop III phase terminates, which in turn depends on whether $\mathcal{T}_c$ or $\mathcal{E}_c$ is in operation. 

\subsection{Spectra of high-$z$ galaxies} \label{sec:spectra}

\begin{figure*}
 \centering
 \includegraphics[width=0.98\textwidth]{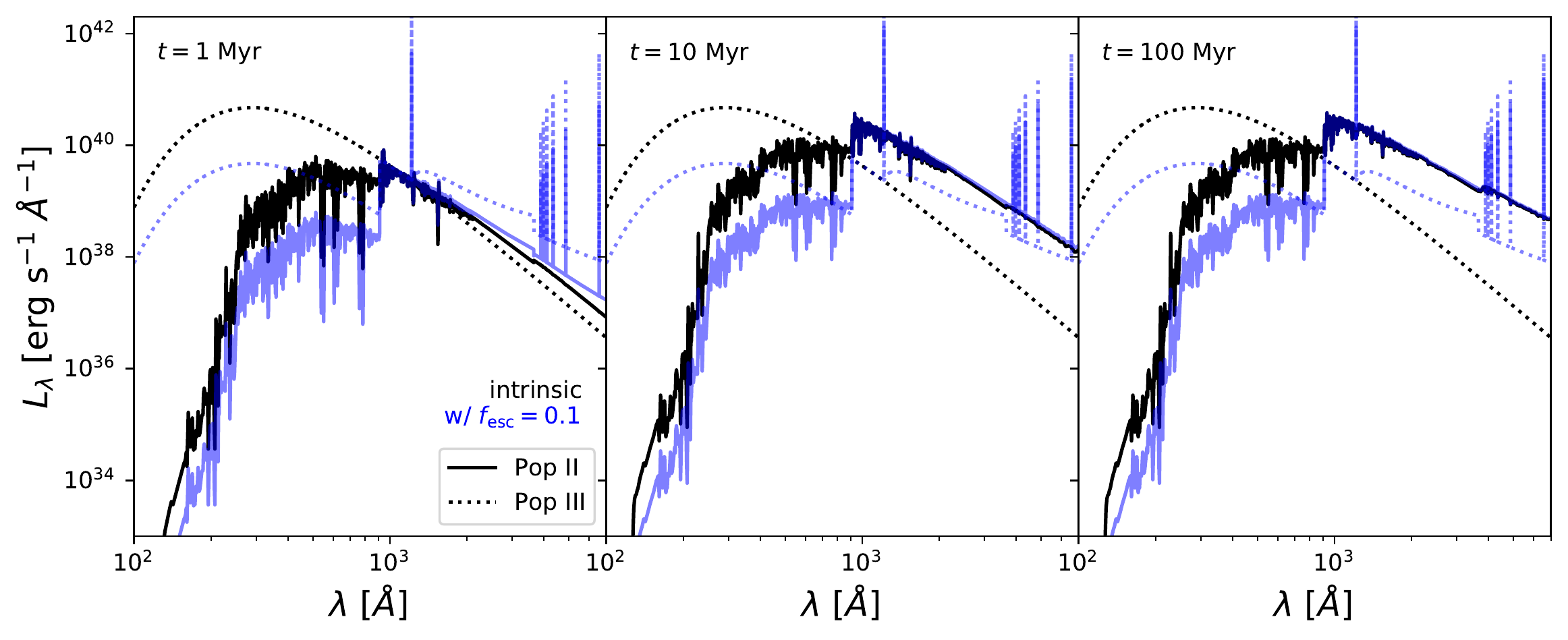}
 \caption{Example spectra of stellar populations employed in this work. In each panel, black curves show the intrinsic Pop~II (solid) and Pop~III (dotted) stellar continuum. For Pop~II, we show models that assume a constant SFR of $1\,M_{\odot}\,\mathrm{yr}^{-1}$ with ages of 1, 10, and 100 Myr (left to right). Pop~III models are the same in each panel, and assume a single star with ionizing luminosity of $10^{48}\,\mathrm{photons}\,\mathrm{s}^{-1}$. Blue lines show the nebular continuum and nebular line emission (see \S\ref{sec:model_mean_la}-\ref{sec:ff_fb}), powered by the absorption of Lyman continuum photons assuming an escape fraction of 10\%. We adopt the $t=100$ Myr models (right-most panel) throughout, a timescale on which the rest-UV spectrum will asymptote to a constant level. The early time evolution is included to demonstrate the nebular continuum treatment.}
 \label{fig:seds}
\end{figure*}

In this section, we introduce our approach to modelling the spectral energy distribution (SED) of high-$z$ galaxies. An illustrative example is shown first in Fig. \ref{fig:seds}, which includes Pop~II and Pop~III spectra, with and without the additional contribution from nebular emission. Each component of the SED is described in more detail in \S\ref{sec:stellar_cont}--\ref{sec:ff_fb}. We note that for the NIRB contribution from nebular line emission we only include hydrogen lines like Ly$\alpha$, the strongest emission line from high-$z$ galaxies in the near-infrared, even though lines such as the \ion{He}{ii} $\lambda$1640 line (for Pop~III stars) could also be interesting --- in the sense of both their contributions to the NIRB and their spatial fluctuations that can be studied in the line-intensity mapping regime. In the following subsections, we specify the individual components of the NIRB according to how they are implemented in \textsc{ares}\footnote{\href{https://github.com/mirochaj/ares}{https://github.com/mirochaj/ares}} \citep{Mirocha_2014}, which was used to conduct all the calculations in this work. 

\subsubsection{Direct stellar emission} \label{sec:stellar_cont}
The direct stellar emission from the surfaces of Pop II and Pop III stars is the foundation upon which the full SED of high-$z$ galaxies is built in our models. It depends in general on the stellar IMF, metallicity, and assumed star formation history of galaxies. For the SED of Pop II stars, we adopt the single-star models calculated with the stellar population synthesis (SPS) code \textsc{bpass} v1.0 \citep{Eldridge2009}, which assume a Chabrier IMF \citep{Chabrier_2003} and a metallicity of $Z=0.02$\footnote{While it is plausible to assume sub-solar metallicity for galaxies during and before reionization given the rate of metal enrichment expected \cite[][]{Furlanetto_2017}, the exact value of $Z$ is highly uncertain and lowering it by 1 or 2 dex does not change our results qualitatively.} in the default case. As is common in many semi-empirical models, we further assume a constant star formation history, for which the rest-UV spectrum evolves little after $\sim 100$ Myr. We therefore adopt 100 Myr as the fiducial stellar population age, as in \citet{Mirocha_2017, Mirocha_2018}, which is a reasonable assumption for high-$z$ galaxies with high specific star formation rates (sSFRs) of the order $10\ \mathrm{Gyr}^{-1}$ \cite[e.g.,][]{Stark_2013}. For Pop III stars, the SED is assumed to be a $10^5$ K blackbody for simplicity, which is appropriate for stars with masses $\gtrsim 100 \ M_{\odot}$ \citep[e.g.,][]{Tumlinson2000,Schaerer_2002}. We further assume that Pop~III stars form in isolation, one after the next, which results in a time-independent SED.

\subsubsection{Ly$\alpha$ emission} \label{sec:model_mean_la}
The full spectrum of a galaxy must also account for reprocessed emission originating in galactic HII regions. The strongest emission line is Ly$\alpha$ --- because Ly$\alpha$ emission is mostly due to the recombination of ionized hydrogen, a simple model for its line luminosity can be derived assuming ionization equilibrium and case-B recombination. Specifically, the photoionization equilibrium is described by defining a volume $V_{\rm S}$ within which the ionization rate equals the rate of recombination
\begin{equation}
\alpha_{\rm B} n^{\rm neb}_e n^{\rm neb}_{\rm H\,II} V_{\rm S} = Q(\mathrm{H})~,
\label{eq:QH}
\end{equation}
where $\alpha_{\rm B} = \alpha^{\rm eff}_{\rm 2 ^2P} + \alpha^{\rm eff}_{\rm 2 ^2S}$ is the total case-B recombination coefficient as the sum of effective recombination coefficients to the $\rm 2 ^2P$ and $\rm 2 ^2S$ states, and $Q(\mathrm{H})$ is the photoionization rate in $\rm s^{-1}$. It is important to note that, in previous models of the NIRB, an additional factor $(1 - f_{\rm esc})$ is often multiplied to $Q_{\rm H}$. It is intended to roughly account for the fraction of ionizing photons actually leaking into the intergalactic medium (IGM), and therefore not contributing to the absorption and recombination processes that source the nebular emission. We have chosen not to take this simple approximation in our model, but to physically connect $f_{\rm esc}$ with the profile of ionizing radiation instead (see Section~\ref{sec:nirb_fluctuation}). The Ly$\alpha$ emission ($\rm 2 ^2P \rightarrow 1 ^2S$) is associated with the recombination of ionized hydrogen to the $\rm 2 ^2P$ state, so its line luminosity can be written as
\begin{equation}
l^{\rm Ly\alpha} = h \nu_{\rm Ly\alpha} \alpha^{\rm eff}_{\rm 2 ^2P} n^{\rm neb}_e n^{\rm neb}_{\rm H\,II} V_{\rm S} = \frac{Q(\mathrm{H}) h \nu_{\rm Ly\alpha} \alpha^{\rm eff}_{\rm 2 ^2P}}{\alpha_{\rm B}}~,
\label{eq:La_caseB}
\end{equation}
or in the volume emissivity $\epsilon_{\nu}^{\rm Ly\alpha}$
\begin{equation}
\epsilon_{\nu}^{\rm Ly\alpha} V_{\rm S} = Q(\mathrm{H}) f_{\rm Ly\alpha} h \nu_{\rm Ly\alpha} \phi(\nu - \nu_{\rm Ly\alpha})~, 
\label{eq:emissivity}
\end{equation}
where $f_{\rm Ly\alpha} = \alpha^{\rm eff}_{\rm 2 ^2P} / \alpha_{\rm B} \approx 2/3$ is the fraction of recombinations ending up as Ly$\alpha$ radiation and $\phi(\nu - \nu_{\rm Ly\alpha})$ is the line profile, which we assume to be a delta function in our model. 

Now, with $\epsilon_b$ being the number of ionizing photons emitted per stellar baryon, which we derive from the stellar spectrum generated with \textsc{bpass} (see \S\ref{sec:stellar_cont}), we can write
\begin{equation}
Q(\mathrm{H}) \approx \epsilon_b \dot{\rho}_{*} V_{\rm S} / m_p~, 
\end{equation}
where $m_p$ is the mass of the proton. The volume emissivity of Ly$\alpha$ photons is then
\begin{equation}
\bar{\epsilon}_\nu^{\rm Ly\alpha} \mathrm d \nu = \frac{\dot{\rho}_{*}}{m_p} f_{\rm Ly\alpha} \epsilon_b h\nu_{\rm Ly\alpha} \phi(\nu - \nu_{\rm Ly\alpha}) d \nu~. 
\end{equation}

It is also important to note that the above calculations assume Ly$\alpha$ emission is completely described by the case-B recombination of hydrogen, which only accounts for the photoionization from the ground state. In practice, though, additional effects such as collisional excitation and ionization may cause significant departures from the case-B assumption. These effects have been found to be particularly substantial for metal-free stars, which typically have much harder spectra than metal-enriched stars (see e.g., \citealt{Raiter_2010} and \citealt{MDF_2016} for details). Due to the deficit of cooling channels, low-metallicity nebulae can have efficient collisional effects that induce collisional excitation/ionization and ionization from excited levels\footnote{\citet{MDF_2016} find the column density and optical depth of hydrogen atoms in the first excited state to be very small in their photoionization simulations using \textsc{Cloudy} \citep{Ferland_2013}, meaning that the photoionization from $n=2$ is likely inconsequential for the boosting.}, which all lead to a higher Ly$\alpha$ luminosity than expected under the case-B assumption. This enhancement is found to scale with the mean energy of ionizing photons. Meanwhile, density effects can mix $\rm 2 ^2S$ and $\rm 2 ^2P$ states, thus altering the relative importance of $\rm Ly\alpha$ and two-photon emission. This is determined simply by $\alpha^{\rm eff}_{\rm 2 ^2P}$ and $\alpha^{\rm eff}_{\rm 2 ^2S}$ in the low-density limit. When density effects are nontrivial as $n_e$ becomes comparable to the critical density $n_{e, \rm crit}$ (at which $\rm 2 ^2S \rightarrow 2 ^2P$ transition rate equals the radiative decay rate), collisions may de-populate the $\rm 2 ^2S$ state of hydrogen before spontaneous decay occurs. In this case, $\rm Ly\alpha$ is further enhanced at the expense of two-photon emission. 

For simplicity, in our model we introduce an ad hoc correction factor $\mathcal{D}_\mathrm{B}$ to account for the net boosting effect of Ly$\alpha$ emission from Pop~III star-forming galaxies. Throughout our calculations, we use a fiducial value of $\mathcal{D}_\mathrm{B}=2$ for Pop~III stars, a typical value for very massive Pop~III stars considered in this work, and $\mathcal{D}_\mathrm{B}=1$ for Pop~II stars. The volume emissivity after correcting for case-B departures is then
\begin{equation}
\bar{\epsilon}_\nu^{\rm Ly\alpha} d \nu = \frac{\dot{\rho}_{*}(z)}{m_p} \epsilon_b h \nu_{\rm Ly\alpha} \mathcal{D}_\mathrm{B} \phi(\nu - \nu_{\rm Ly\alpha}) d \nu~. 
\end{equation}
We also note that, by default, our nebular line model also includes Balmer series lines, using line intensity values from Table 4.2 of \citet{OF_2006AGN3}.

\subsubsection{Two-photon emission} \label{sec:model_mean_tp}

For two-photon emission ($\rm 2 ^2S \rightarrow \rm 1 ^2S$), the probability of transition producing \textit{one} photon with frequency in range $\mathrm d x = \mathrm d \nu / \nu_{\rm Ly\alpha}$ can be modelled as \cite[][]{FK_2006}
\begin{equation}
P(x') = 1.307 - 2.627x'^2 + 2.563x'^4 - 51.69x'^6~, 
\end{equation}
where $x'=x-0.5$. Note that $P(x')$ is symmetric around $x=0.5$ as required by energy conservation and is normalized such that $\int_0^1 P(x) d x = 1$. By analogy to $\rm Ly\alpha$ emission, the two-photon volume emissivity under the case-B assumption can be written as
\begin{equation}
\bar{\epsilon}_\nu^{2\gamma} d \nu = \frac{\dot{\rho}_{*}(z)}{m_p} (1-f_{\rm Ly\alpha}) \epsilon_b \frac{2 h \nu}{\nu_{\rm Ly\alpha}} P(\nu/\nu_{\rm Ly\alpha}) d \nu~. 
\end{equation}

\subsubsection{Free-free \& free-bound emission} \label{sec:ff_fb}

The free-free and free-bound (recombination to different $n$ levels of hydrogen) emission also contribute to the nebular continuum. The specific luminosity and the volume emissivity are related by
\begin{equation}
l_\nu = \frac{\epsilon_\nu Q_{\rm H}}{n_e n_p \alpha_{\rm B}}~,
\label{eq:Lnu}
\end{equation}
where $\alpha_{\rm B}$ as a function of gas temperature $T_g$ is given by
\begin{equation}
\alpha_{\rm B} = \frac{2.06 \times 10^{-11}}{T_g^{1/2}} \phi_2(T_g) \sim \frac{2.06 \times 10^{-11}}{T_g^{1/2}} \, \mathrm{cm^3\,s^{-1}}~, 
\end{equation}
where $\phi_2(T_g)$ is a dimensionless function of gas temperature that is of order unity for a typical temperature of \ion{H}{ii} regions $T_g \approx 2\times10^4$\,K. We take the following expression given by \citet[][]{DS_2003} for the volume emissivity including both free-free and free-bound emission 
\begin{equation}
\epsilon_\nu^{\rm free} = 4 \pi n_e n_p \gamma_\mathrm{c}(\nu) \frac{e^{-h\nu/kT_g}}{T_g^{1/2}} \, \rm{erg\,cm^{-3}\,s^{-1}\,Hz^{-1}}~, 
\end{equation}
where a continuous emission coefficient, $\gamma_\mathrm{c}(\nu)$, in units of $\mathrm{cm^3\,erg\,s^{-1}\,Hz^{-1}}$ is introduced to describe the strengths of free-free and free-bound emission. Values of $\gamma_\mathrm{c}$ as a function of frequency are taken from Table~1 of \citet{Ferland1980}, which yield a nebular emission spectrum in good agreement with the reprocessed continuum predicted by photoionization simulations. We can then write the emissivity as
\begin{align}
\bar{\epsilon}_\nu^{\rm free} d \nu & = \frac{4\pi}{2.06 \times 10^{-11}} \frac{\dot{\rho}_{*}(z)}{m_p} \epsilon_b e^{-h\nu/kT} \gamma_c(\nu) d \nu~. 
\label{eq:fffb_emiss}
\end{align}

Note that the volume emissivities shown above with an overbar can be considered as the first moment of luminosity, namely averaging the luminosity per halo over the halo mass function
\begin{equation}
\bar{\epsilon}_{\nu}^i(z) = \int n(M_h) l_{\nu}^i(M_h, z) d M_h~,
\end{equation}
where $l_{\nu}^i(M_h, z)$ is the specific luminosity of component $i$ as a function of halo mass and redshift, which can be obtained by simply replacing the SFRD, $\dot{\rho}_{*}$, in equation~ \ref{eq:fffb_emiss} with the star formation rate, $\dot{M}_{*}$.

\subsection{Mean NIRB intensity}

For a given source population, the mean intensity at an observed frequency $\nu_0$ of the NIRB can be described by evolving the volume emissivity through cosmic time (i.e., the solution to the cosmological radiative transfer equation)
\begin{equation}
J_{\nu_0}(z) = \frac{1}{4\pi} \int_{z_0}^{z} d z' \frac{d \ell}{d z'} \frac{(1+z_0)^3}{(1+z')^3} \bar{\epsilon}^{\rm prop}_{\nu'}(z') e^{-\tau_{\rm HI}(\nu, z_0, z')}~,
\label{eq:Jnu0}
\end{equation}
where $d \ell / d z' = c / [H(z') (1+z')]$ is the proper line element and $\nu' = \nu_0(1+z')/(1+z_0)$. For $z_0 = 0$, the average, comoving volume emissivity is related to the proper volume emissivity by $\bar{\epsilon}_{\nu}(z) = \bar{\epsilon}^{\rm prop}_{\nu}(z)/(1+z)^3$. If one assumes the IGM is generally transparent to NIRB photons from high redshifts, then the mean intensity can be simplified to~\cite[e.g.,][]{Fernandez_2010, Yang_2015}
\begin{equation}
J_{\nu} \equiv \bar{I}_{\nu} = \frac{c}{4\pi} \int d z \frac{\bar{\epsilon}_{\nu'}(z)}{H(z)(1+z)}~,
\label{eq:mean_Inu}
\end{equation}
or the per logarithmic frequency form~\cite[e.g.,][]{Cooray_2012}, 
\begin{equation}
\nu \bar{I}_{\nu} = \frac{c}{4\pi} \int d z \frac{\nu' \bar{\epsilon}_{\nu'}(z)}{H(z)(1+z)^2}~. 
\end{equation}
However, the IGM absorption may not be negligible for certain NIRB components, such as the highly resonant Ly$\alpha$ line, in which case the radiative transfer equation must be solved in detail. To approximate the attenuation by a clumpy distribution of intergalactic \ion{H}{i} clouds, we adopt the IGM opacity model from \citet{Madau_1995}. In \textsc{ares}, equation (\ref{eq:mean_Inu}) is solved numerically following the algorithm introduced in \citet{Haardt1996}.

\subsection{NIRB fluctuations}
\label{sec:nirb_fluctuation}

Using the halo model established by \citet{CS_2002}, we can express the three-dimensional (3D), spherically-averaged power spectrum of the NIRB anisotropy associated with high-$z$ galaxies as a sum of three terms
\begin{equation}
P_{\rm NIR}(k, z) = P_{\rm 2h}(k, z) + P_{\rm 1h}(k, z) + P_{\rm shot}(z)~, 
\label{eq:PNIR_3D}
\end{equation}
where each term is composed of direct stellar emission and/or nebular emission. In our model, we divide the emission from a galaxy into two components: (1) a discrete, point-source-like component sourced by direct stellar emission and contributing to the two-halo and shot-noise terms, and (2) a continuous, spatially-extended component sourced by nebular emission from the absorption of ionizing photons in the circumgalactic medium (CGM) or IGM by neutral gas and contributing to the two-halo and one-halo terms. 

Specifically, the two-halo term is proportional to the power spectrum of the underlying dark matter density field
\begin{equation}
P_{\rm 2h}(k) = \left[ \int n(M_h) b(M_h) \sum_i l_\nu^i(M_h) u_i(k|M_h) d M_h \right]^2 P_{\rm \delta \delta}(k),
\end{equation}
where the summation is over the stellar and nebular components of galactic emission and $u(k)$ is the normalized Fourier transform of the halo flux profile. $P_{\rm \delta \delta}$ is the dark matter power spectrum obtained from \textsc{CAMB} \citep{LCL_2000}. We take $u_*(k)=1$ for the halo luminosity of direct stellar emission ($l^{*}_{\nu}$) and derive the functional form $u_{\rm n}(k)$ for the halo luminosity of nebular emission ($l_\nu^\mathrm{Ly\alpha}$, $l_\nu^\mathrm{2\gamma}$, $l_\nu^\mathrm{ff+fb}$) using the profile of ionizing flux emitted from the galaxy. Because the one-halo term is only sourced by nebular emission, it can be expressed as
\begin{equation}
P_{\rm 1h}(k) = \int n(M_h) \left[ \sum_j l_{\nu}^{j}(M_h) u_{\rm n}(k|M_h) \right]^2 d M_h~,
\end{equation}
where the summation is over the different types of nebular emission described in \S\ref{sec:model_mean_la}--\ref{sec:ff_fb}. Finally, the scale-independent shot-noise term is solely contributed by direct stellar emission, namely
\begin{equation}
P_{\rm shot} = \int n(M_h) \left[ l^{*}_{\nu}(M_h) \right]^2 d M_h~.
\label{eq:pshot}
\end{equation}
For simplicity, we ignore the stochasticity in luminosity--halo mass relations for the ensemble of galaxies. Its effect on the shape of $P_{\rm NIR}(k)$ may be quantified by assuming a probability distribution function \cite[e.g.,][]{Sun_2019}, but is likely subdominant to (and degenerate with) the systematic uncertainties associated with the relations themselves.

\subsubsection{The radial profile of nebular emission}

We stress that in our model, the nebular emission is assumed to be smooth and thus contributes to $P_{\rm 2h}$ and $P_{\rm 1h}$ only. In addition, rather than treating $f_{\rm esc}$ as a completely free parameter, we determine its value from the profile of ionizing flux, which in turn depends on the neutral gas distribution surrounding galaxies. This effectively renders $f_{\rm esc}$ and the shape of the one-halo term, which is captured by $u_{\rm n}(k|M)$, dependent on each other. 

To derive $u_{\rm n}(k|M)$, we consider the scenario in which ionizing photons are radiated away from the centre of galaxy under the influence of neutral gas distribution in the CGM. While ionizing photons escaped into the IGM can also in principle induce large-scale fluctuations of the types of nebular emission considered in this work, especially Ly$\alpha$, their strengths are found to be subdominant to the emission close to galaxies \cite[e.g.,][]{Cooray_2012}. For the CGM, since a substantial overdensity of neutral hydrogen exists in the circumgalactic environment in the high-redshift universe, the extended Ly$\alpha$ (and other nebular) emission is primarily driven by the luminosity of the ionizing source and the distribution of neutral gas clumps surrounding it. Here we only provide a brief description of the neutral gas distribution models adopted and refer interested readers to \cite{MD_2016} and \cite{MasRibas_2017} for further details. For the Ly$\alpha$ flux resulting from the fluorescent effect in the CGM, the radial profile at a proper distance $r$ scales as
\begin{equation}
\mathrm d F_{\rm Ly\alpha}(r) \propto - r^{-2} f_{\rm c}(r) f_{\rm esc}(r) d r~,
\end{equation}
where $r^{-2}$ describes the inverse-square dimming and $f_{\rm esc}(r) = \exp{-[\int_0^r f_{\rm c}(r') d r']}$ represents the fraction of ionizing photons successfully escaped from the ionizing source at distance $r$. $f_{\rm c}(r)$ is the \textit{differential}, radial covering fraction of \ion{H}{i} clumps, whose line-of-sight integral gives the total number of clumps along a sight line, analogous to the number of mean free path lengths. The product $f_{\rm c}f_{\rm esc}$ can be interpreted as the chance that an ionizing photon gets absorbed by a clump of \ion{H}{i} cloud and thus gives rise to a Ly$\alpha$ photon. The resulting flux profile can then be expressed as
\begin{equation}
F_{\rm Ly\alpha}(r) \propto \int_{r}^{\infty} r'^{-2} f_{\rm c}(r') f_{\rm esc}(r') d r'~,
\end{equation}
given the boundary condition $F_{\rm Ly\alpha} = 0$ as $r \rightarrow \infty$.

\begin{figure}
 \centering
 \includegraphics[width=0.48\textwidth]{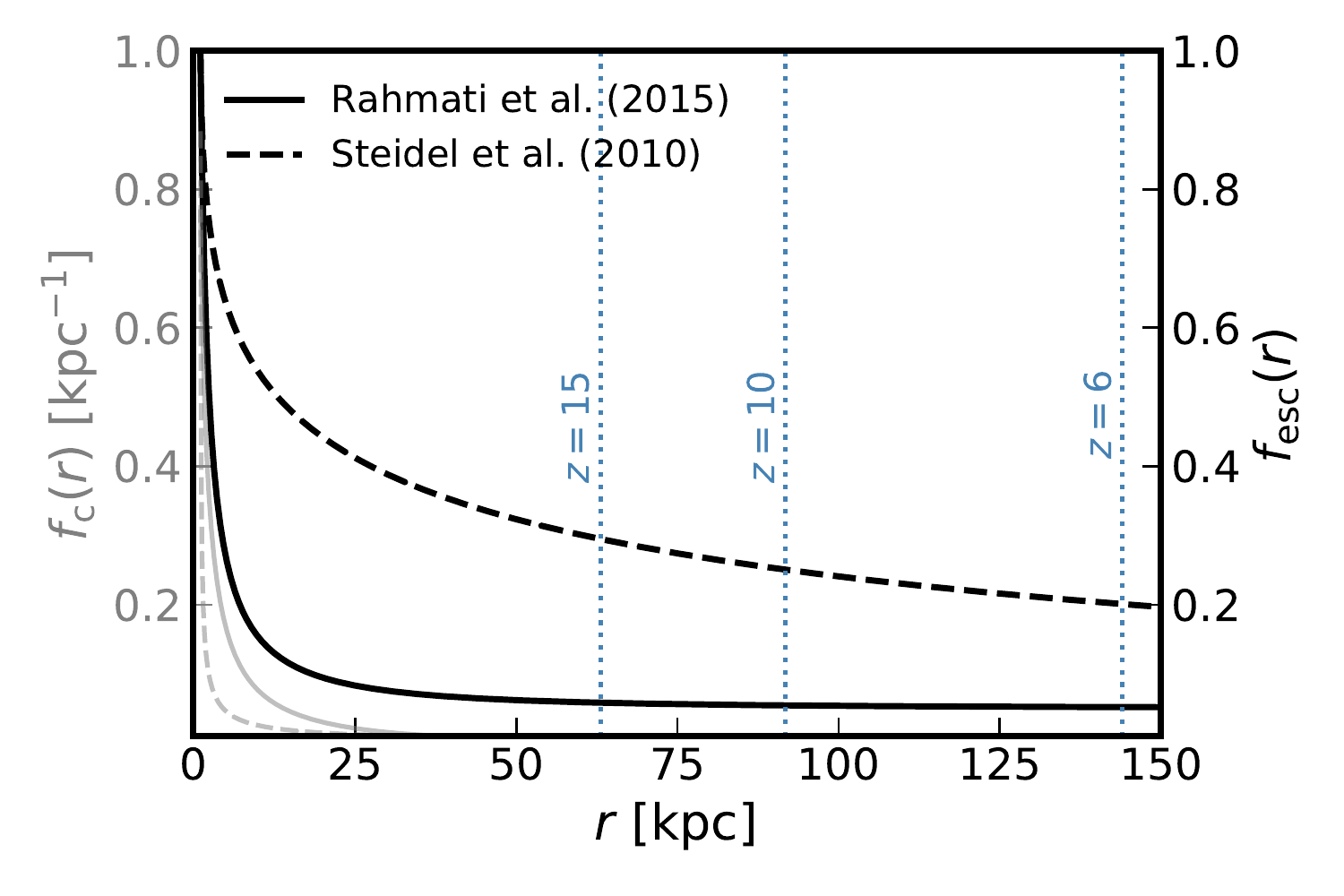}
 \caption{The radial profiles of the \ion{H}{I} covering fraction $f_{\rm esc}$ (grey, left axis) and the escape fraction of ionizing photons $f_{\rm esc}$ (black, right axis) as functions of the radial distance $r$ away from the galaxy, derived from two CGM models by \citet{Rahmati_2015} and \citet{Steidel_2010}. The virial radius of a $10^{14}\,M_\odot$ halo, which defines an upper bound on the scale relevant to ionizing photons escaping into the IGM, is quoted at $z=6$, 10, and 15 (dotted vertical lines).}
 \label{fig:fprofile}
\end{figure}

Various CGM models have been proposed for high-$z$ galaxies, from which the \ion{H}{i} covering fraction $f_{\rm c}(r)$ can be obtained. However, due to the paucity of observational constraints especially in the pre-reionization era, it is impractical to robustly determine which one best describes the nebular emission profile of high-$z$ galaxies relevant to our model. As a result, we follow \citet{MasRibas_2017} and consider two CGM models that predict distinct \ion{H}{i} spatial distributions surrounding galaxies, leading to high and low escape fractions of ionizing photons, respectively. We caution that the two profiles are explored here only to demonstrate the connection between $f_{\rm esc}(r)$ and small-scale fluctuations. Exact escape fractions they imply are assessed with other observational constraints, such as the CMB optical depth, and therefore some tension may exist for a subset of our Pop~III models. We will revisit this point in Section~\ref{sec:imply-reionization}.

The low-leakage model is based on the fitting formula (see equation~17 of \citealt{MD_2016}) for the area covering fraction of Lyman limit systems (LLSs), $\mathcal{F}_{\rm LLS}$, inferred from the EAGLE simulation \cite[][]{Rahmati_2015}. It has been successfully applied to reproduce the observed stacked profile of extended Ly$\alpha$ emission from Lyman-alpha emitters (LAEs) out to $z = 6.6$. Specifically, the radial covering fraction $f_{\rm c}$ is related to the area covering fraction $\mathcal{F}_{\rm LLS}(b)$, defined for a total area of $2\pi b d b$ at the impact parameter $b$, by an inverse Abel transformation
\begin{equation}
f_{\rm c}(r) = - \frac{1}{\pi} \int_r^\infty \frac{d N_{\rm clump}}{d y} \frac{d y}{\sqrt{y^2 - r^2}}~,
\end{equation}
where the number of gas clumps encountered is given by $N_{\rm clump}(b) = - \ln\left[1 - \mathcal{F}_{\rm LLS}(b)\right]$. The high-leakage model is proposed by \citet[][]{Steidel_2010} to provide a simple explanation to interstellar absorption lines and Ly$\alpha$ emission in the observed far-UV spectra of Lyman break galaxies (LBGs) at $z\lesssim3$. It describes a clumpy outflow consisting of cold \ion{H}{i} clumps embedded within a hot medium accelerating radially outward from the galaxy. The radial covering fraction $f_{\rm c}$ in this case can be written as \citep{DK_2012}
\begin{equation}
f_{\rm c}(r) = n_{\rm c}(r) \pi R^2_{\rm c}~,
\end{equation}
where $n_{\rm c}(r)$ is the number density of the \ion{H}{i} clumps that is inversely proportional to their radial velocity $v(r)$ determined from the observed spectra, and the clump radius $R_{\rm c} \propto r^{-2/3}$ under pressure equilibrium. 

Fig.~\ref{fig:fprofile} shows a comparison between radial profiles of $f_{\rm c}$ and $f_{\rm esc}$ in the two CGM models considered. The higher \ion{H}{i} covering fraction in the \citet{Rahmati_2015} model results in an $f_{\rm esc}$ profile which declines more rapidly with $r$ than that from the \citet{Steidel_2010} model. Given the potentially large uncertainties associated with the exact mapping between the $f_{\rm esc}$ profile and the average escape fraction $\bar{f}_{\rm esc}$ that matters for reionization, we refrain from defining $\bar{f}_{\rm esc}$ at the virial radius of a halo that hosts a typical EoR galaxy, as done by \citet{MasRibas_2017}. Instead, we quote the value of $f_{\rm esc}$ as predicted by the two CGM models at a proper distance $r = 150\,\mathrm{kpc}$, sufficiently large compared to the virial radii of the largest relevant haloes ($10^{14}\,M_\odot$) as shown by the vertical dotted lines in Fig.~\ref{fig:fprofile}. This allows us to effectively define \textit{lower bounds} on the average escape fraction $\bar{f}_{\rm esc}=0.05$ and $0.2$ corresponding to the \citet{Rahmati_2015} and \citet{Steidel_2010} models, respectively, which in turn set \textit{upper bounds} on the nebular emission signal allowed in the two cases. We note, nevertheless, that both CGM models predict only modest evolution of $f_{\rm esc}(r)$ beyond a few tens kpc --- the size range of more typical haloes hosting ionizing sources. The exact choice of $\bar{f}_{\rm esc}$ value is thus expected to have only a small impact on the NIRB signal predicted, whereas the corresponding reionization history is more sensitive to this choice, as will be discussed in Section~\ref{sec:imply-reionization}. To simplify the notation, in what follows we will drop the bar and use $f_\mathrm{esc}$ to denote the lower bound on $\bar{f}_\mathrm{esc}$ inferred from the CGM model chosen. As summarized in Table~\ref{tb:model_params}, in our models we set $f^\mathrm{III}_\mathrm{esc} = 0.05$ or $0.2$ for Pop~III stars according to the two CGM models, whereas for Pop~II stars we adopt an intermediate profile that yields $f^\mathrm{II}_\mathrm{esc} = 0.1$. With reasonable faint-end extrapolations as in our model, an escape fraction of 10\% is proven to yield a reionization history consistent with current observations without the presence of unknown source populations like Pop~III stars. 

\subsubsection{The angular power spectrum}

Following \cite{Fernandez_2010} and \cite{LF_2013}, we can derive the angular power spectrum from the 3D power spectrum. With an observed frequency $\nu$, equation~(\ref{eq:mean_Inu}) gives the NIRB intensity, which can be expressed as a function of direction on the sky $\hat{\mathbf{n}}$
\begin{equation}
I_\nu(\hat{\mathbf{n}}) = \frac{c}{4\pi} \int_{z_{\rm min}}^{z_{\rm max}} \frac{\epsilon_{\nu'}[z, \hat{\mathbf{n}}r(z)]}{H(z)(1+z)} d z~, 
\label{eq:specific_mean_intensity}
\end{equation}
where $\nu'=(1+z)\nu$ and $r(z)$ is the comoving radial distance out to a redshift $z$. Spherical harmonics decomposing $I_\nu(\hat{\mathbf{n}})$ gives
\begin{equation}
I_\nu(\hat{\mathbf{n}}) = \sum_{\ell, m} a_{\ell m} Y_\ell^m(\hat{\mathbf{n}})~,
\end{equation}
with the coefficient 
\begin{equation}
a_{\ell m} = \frac{c}{4\pi} \int \frac{d z \int d \hat{\mathbf{n}} \int \frac{d^3 \mathbf{k}}{(2\pi)^3} \epsilon_{\nu'}(z, \mathbf{k}) e^{-i\mathbf{k} \cdot \hat{\mathbf{n}}r(z)} Y^*_{\ell m}(\hat{\mathbf{n}})}{H(z)(1+z)}~.
\end{equation}
Using Rayleigh's formula for $e^{-i\mathbf{k} \cdot \hat{\mathbf{n}}r(z)}$, we have
\begin{equation}
a_{\ell m} = \int \frac{c (-1)^\ell d z }{H(z)(1+z)} \int \frac{d^3 \mathbf{k}}{(2\pi)^3} \tilde{\epsilon}_{\nu'}(z, \mathbf{k}) j_\ell[kr(z)] Y^*_{\ell m}(\hat{\mathbf{k}})~.
\end{equation}
The angular power spectrum is consequently defined as the ensemble average $C_\ell = \langle |a_{\ell m}|^2 \rangle$. For a pair of observed frequencies $\nu_1$ and $\nu_2$, it can be written as (assuming Limber's approximation, which is valid for the range of $\ell \gg 1$ considered in this work)
\begin{equation}
C_\ell^{\nu_1 \nu_2} = \ \frac{c}{(4\pi)^2} \int \frac{P^{\nu_1 \nu_2}_{\rm NIR}\left[\nu_1(1+z), \nu_2(1+z), \ell/r(z)\right] d z}{H(z)r^2(z)(1+z)^2}~, 
\label{eq:cl_nn}
\end{equation}
where $P^{\nu_1 \nu_2}_{\rm NIR}$ is the 3D NIRB power spectrum defined in equation~\ref{eq:PNIR_3D}.

Alternatively, a \textit{band-averaged} intensity may be defined, in which case a factor of $(1+z)$ must be introduced to account for the cosmological redshift \citep{Fernandez_2010}. Namely, in contrast to equation~\ref{eq:specific_mean_intensity}, we have
\begin{align}
I(\hat{\mathbf{n}}) & = \frac{1}{\Delta \nu} \int_{\nu_1}^{\nu_2} d \nu I_{\nu}(\hat{\mathbf{n}})\nonumber \\
& = \frac{c}{4\pi \Delta \nu} \int d z \frac{\int_{\nu_1(1+z)}^{\nu_2(1+z)} d \tilde{\nu} \epsilon_{\tilde{\nu}}[z, \hat{\mathbf{n}}r(z)]}{H(z)(1+z)^2} \nonumber \\ 
& = \frac{c}{4\pi \Delta \nu} \int d z \frac{\rho^{\rm em}_{L}[z, \hat{\mathbf{n}}r(z)]}{H(z)(1+z)^2}~,
\end{align}
where $\rho^{\rm em}_{L}$ represents the luminosity density emitted over some frequency band at the corresponding redshift. The band-averaged angular power spectrum is then
\begin{equation}
C_\ell = \frac{c}{(4\pi \Delta \nu)^2} \int \frac{d z}{H(z)r^2(z)(1+z)^4} P_{L}^{\rm NIR}\left[k=\ell/r(z), z\right]~.
\label{eq:cl}
\end{equation}

% -------------------------- S3: Results -------------------------- %

\begin{figure}
 \centering
 \includegraphics[width=0.48\textwidth]{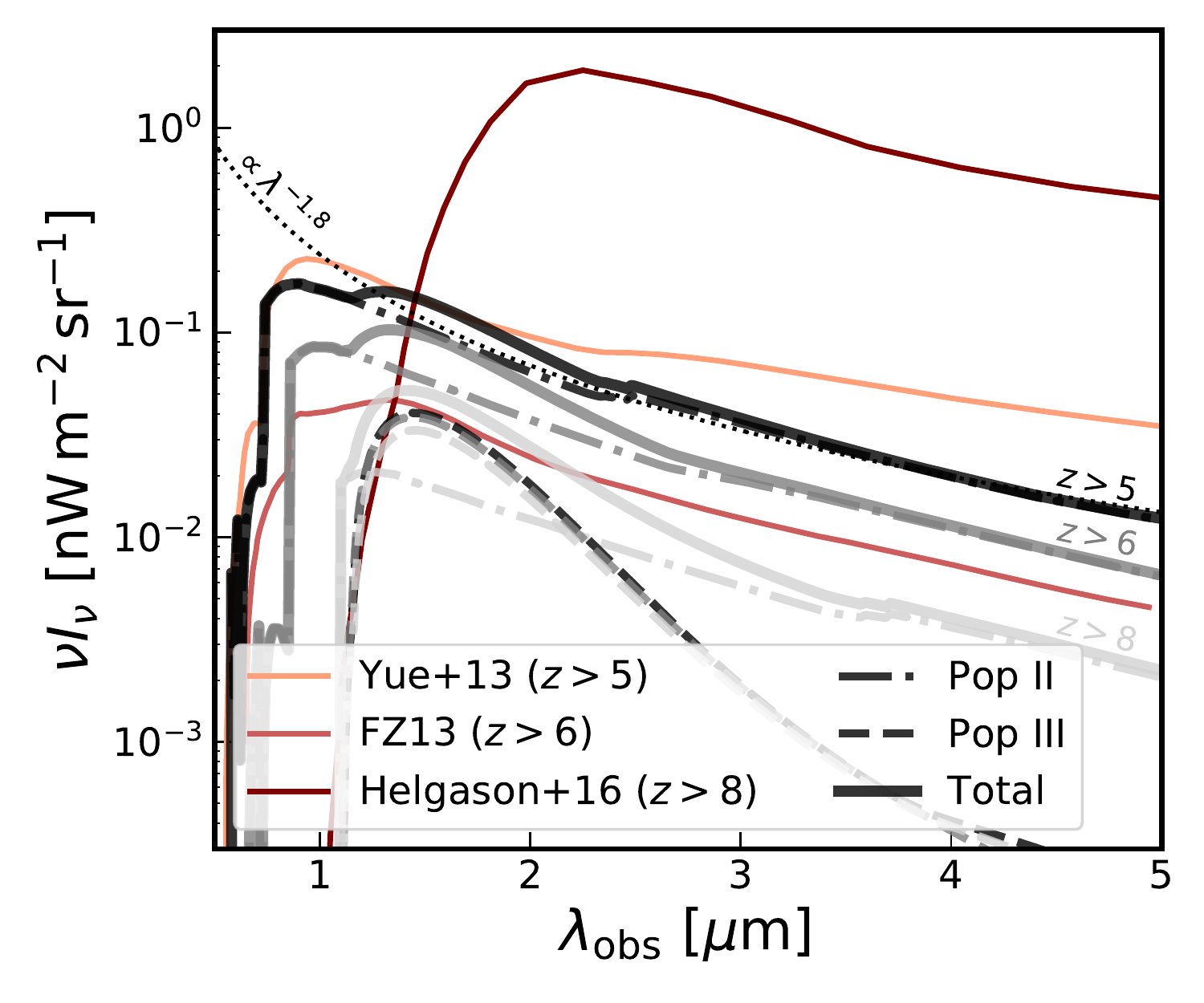}
 \caption{The spectra of NIRB mean intensity $\nu \bar{I}_\nu$ sourced by Pop~II (dash-dotted) and Pop~III (dashed) star-forming galaxies at different redshifts, predicted by Model~IA. The Pop~II contribution can be approximated by $\lambda^{-1.8}$. For comparison, we show in color a few model predictions in the literature that include contributions from both Pop~II and Pop~III stars (\citealt{Yue_2013}; \citetalias{FZ_2013}; \citealt{Helgason_2016}). The impact of the Pop~III to Pop~II transition, which varies significantly among these models, can be seen from the shape and amplitude of NIRB spectrum. A spectral peak redward of 1 micron is characteristic of a significant Ly$\alpha$ contribution to the NIRB intensity due to the efficient formation of massive, Pop~III stars.}
 \label{fig:nuInu}
\end{figure}

\section[]{Results} \label{sec:results}

In this section, we show the high-$z$ NIRB signals sourced by galaxies at $z>5$, with the emphasis on the potential contribution of Pop~III stars. We first present a general picture expected given our reference model which combines a semi-empirical description of the known, Pop~II star-forming galaxies and an optimistic model of Pop~III star formation, characterized by high Pop~III SFR with relatively inefficient chemical feedback (Section~\ref{sec:result:nirb_zgt5}). Then, by exploring a range of plausible Pop~III star formation histories, we focus on how spectral signatures of Pop~III stars on the NIRB connect to their properties (Section~\ref{sec:results:imprints}). Finally, we estimate the sensitivities of two future instruments, SPHEREx and CDIM, to the high-$z$ NIRB signals (Section~\ref{sec:results:sensitivity}). 

\subsection{The NIRB from star-forming galaxies at $z>5$} \label{sec:result:nirb_zgt5}

To provide a general picture of the NIRB signal associated with first galaxies, we define our reference model to be Model~IA, as specified in Table~\ref{tb:model_params}. The SFE of Pop~II stars $f_*$ follows a double power-law in mass fit to the observed galaxy UVLFs over $5<z<10$, and the Pop~III SFRD is tuned such that the total cosmic SFRD roughly matches the maximum-likelihood model from \citet{Robertson_2015} based on the electron scattering optical depth $\tau_e$ of CMB photons from Planck. A set of variations around this baseline case will be considered in the subsections that follow. 

In Fig.~\ref{fig:nuInu}, we show the mean intensity spectra of the NIRB over  $0.75$--$5\,\mu\mathrm{m}$, calculated from Model~IA with different redshift cutoffs. For comparison, results from the literature that account for both Pop~II and Pop~III stars with similar cutoffs are also displayed. The sharp spectral break at the Ly$\alpha$ wavelength redshifted from the cutoff is caused by the IGM attenuation as described by \citet{Madau_1995}, which serves as a characteristic feature that distinguishes the high-$z$ component from low-$z$ ones. From our model, the NIRB spectrum associated with Pop~II stars without being blanketed by \ion{H}{i} blueward of Ly$\alpha$ is predominantly sourced by direct stellar emission, and it can be well described by a power law that scales as $\lambda^{-1.8}$. This roughly agrees with the Pop-II-dominated prediction from \citet{Yue_2013}, who find a slightly shallower slope that might be attributed to different assumptions adopted in the SED modelling and the SFH assumed. Unlike Pop~II stars, massive Pop~III stars contribute to the NIRB mainly through their nebular emission, especially in Ly$\alpha$. The resulting NIRB spectrum therefore has a much stronger wavelength dependence that traces the shape of the Pop~III SFRD. Similar to \citetalias{FZ_2013}, our reference model suggests that strong Ly$\alpha$ emission from Pop~III stars may lead to a spectral ``bump'' in the total NIRB spectrum, which causes an abrupt change of spectral index over $1$--$1.5\,\mu$m. We will discuss the implications of such a Pop~III signature in detail in Section~\ref{sec:results:imprints}. We also compare our Pop~III prediction based on physical arguments of different feedback mechanisms, to an extreme scenario from \citet{Helgason_2016} attempting to explain the entire observed, source-subtracted NIRB fluctuations with the Pop~III contribution. The fact that our reference model, which already makes optimistic assumptions about the efficiency of Pop~III star formation, predicts more than an order of magnitude lower NIRB signal corroborates the finding of \citet{Helgason_2016}. Pop~III stars alone are unlikely to fully account for the observed NIRB excess without violating other observational constraints such as the reionization history --- unless some stringent requirements on the physics of Pop~III stars are met, including their ionizing and metal production efficiencies.  

\begin{figure}
 \centering
 \includegraphics[width=0.48\textwidth]{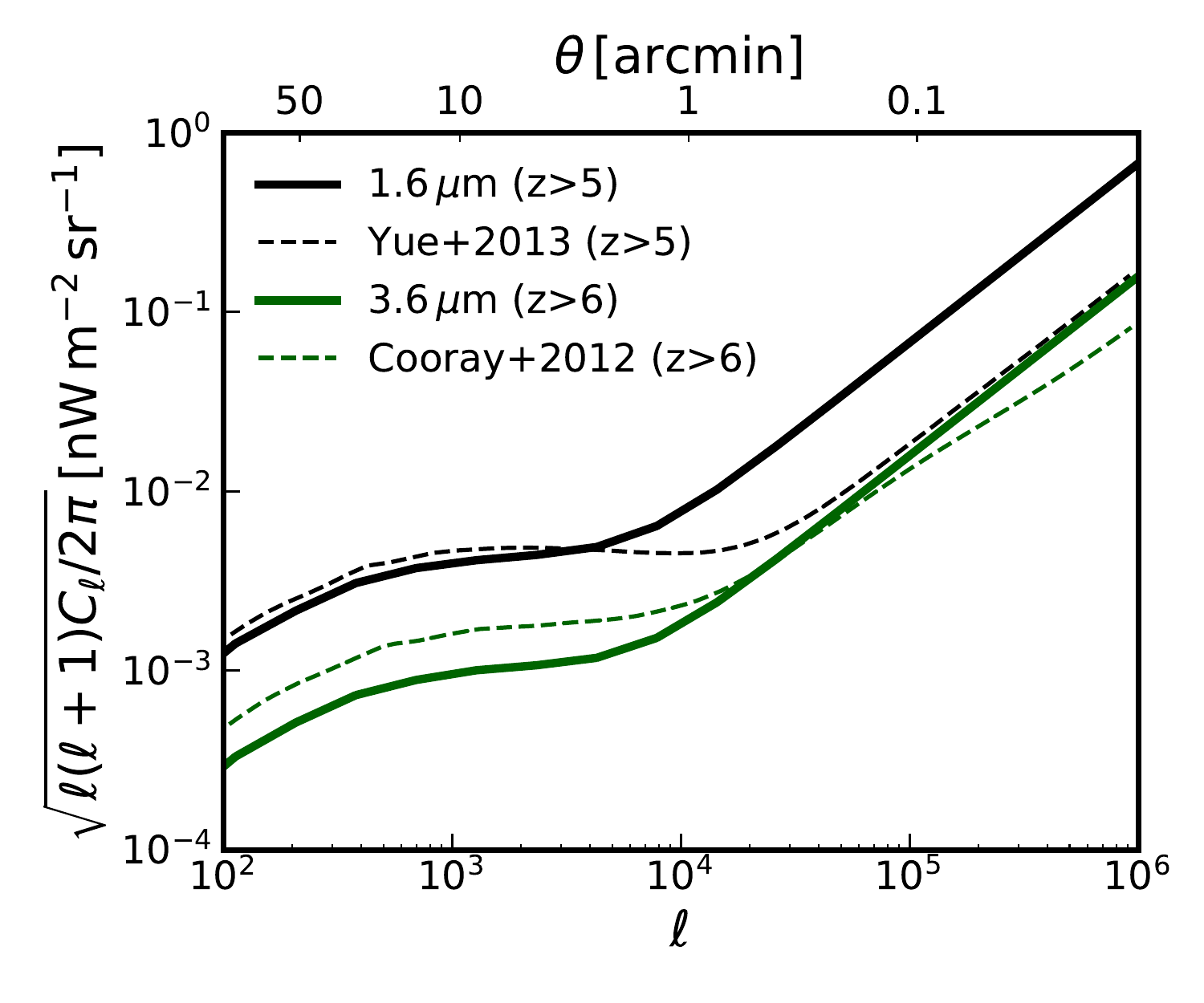}
 \caption{Comparison of the NIRB angular power spectra associated with Pop~II and Pop~III stars over different bands and redshift ranges. As opposed to \citet{Cooray_2012} and \citet{Yue_2013}, our model predicts a higher shot-noise power due to the inefficient star formation in low-mass haloes as described by the mass-dependent $f_*$.}
 \label{fig:dF_comp}
\end{figure}

Fig.~\ref{fig:dF_comp} shows predicted the angular intensity fluctuations $\delta F = \sqrt{\ell(\ell+1)C_\ell/2\pi}$ of the NIRB by our reference model at two wavelengths, $1.6$ and $3.6\,\mu\mathrm{m}$. Compared with predictions at the same wavelengths from \citet{Cooray_2012} and \citet{Yue_2013}, our model produces similar (within a factor of 2) large-scale clustering amplitudes. On small scales, our model predicts significantly higher shot-noise amplitudes. Such a difference in the shape of angular power spectrum, $C_\ell$, underlines the importance of properly accounting for the contribution from the population of faint/low-mass galaxies loosely constrained by observations. While all these models assume that haloes above a mass $M_{\rm min} \sim 10^8\,M_\odot$ can sustain the formation of Pop~II stars (which dominates the total NIRB fluctuations) through efficient atomic cooling of gas, our model allows $f_*$ to evolve strongly with halo mass. As demonstrated in a number of previous works \citep{Moster_2010, Mirocha_2017, Furlanetto_2017}, the observed UVLFs of galaxies at $z>5$ can be well reproduced by $f_*$ as a double power-law in halo mass, consistent with simple stellar and AGN feedback arguments that suppress star formation in low-mass and high-mass haloes, respectively. Consequently, low-mass haloes in our model, though still forming stars at low levels, contribute only marginally to the observed NIRB fluctuations, especially on small scales where the Poissonian distribution of bright sources dominates the fluctuations. The resulting angular power spectrum has a shape different from those predicted by \citet{Cooray_2012} and \citet{Yue_2013}, with fractionally higher shot-noise amplitude. Measuring the full shape of $C_\ell$ from sub-arcminute scales (where the sensitivity to $f_*$ maximizes) to sub-degree scales (where the high-$z$ contribution maximizes) with future NIRB surveys can therefore place interesting integral constraints on the effect of feedback regulation on high-$z$, star-forming galaxies, complementary to measuring the faint-end slope of the galaxy UVLF. 

\subsection{Spectral signatures of first stars on the NIRB} \label{sec:results:imprints}

As shown in Fig.~\ref{fig:nuInu}, a characteristic spectral signature may be left on the NIRB spectrum in the case of efficient formation of massive Pop~III stars. Details of such a feature, however, depend on a variety of factors involving the formation and physical properties of both Pop~II and Pop~III stars. Of particular importance is when and for how long the transition from Pop~III stars to Pop~II stars occurred, which can be characterized by the ratio of their SFRDs, even though stellar physics such as age and the initial mass function (IMF) also matter and therefore serve as potential sources of degeneracy. \citetalias{FZ_2013} studies the NIRB imprints in this context using a simple phenomenological model for the Pop~III to Pop~II transition, without considering detailed physical processes that drive the transition. In this subsection, we investigate the effects of varying the Pop~II and Pop~III SFHs separately on the NIRB signal from high-$z$ galaxies, exploring 
a set of physically-motivated model variations specified in Table~\ref{tb:model_params}. 

\subsubsection{Effects of variations in the Pop~II SFH}
\label{sec:results:imprints:pop2}

\begin{figure}
 \centering
 \includegraphics[width=0.48\textwidth]{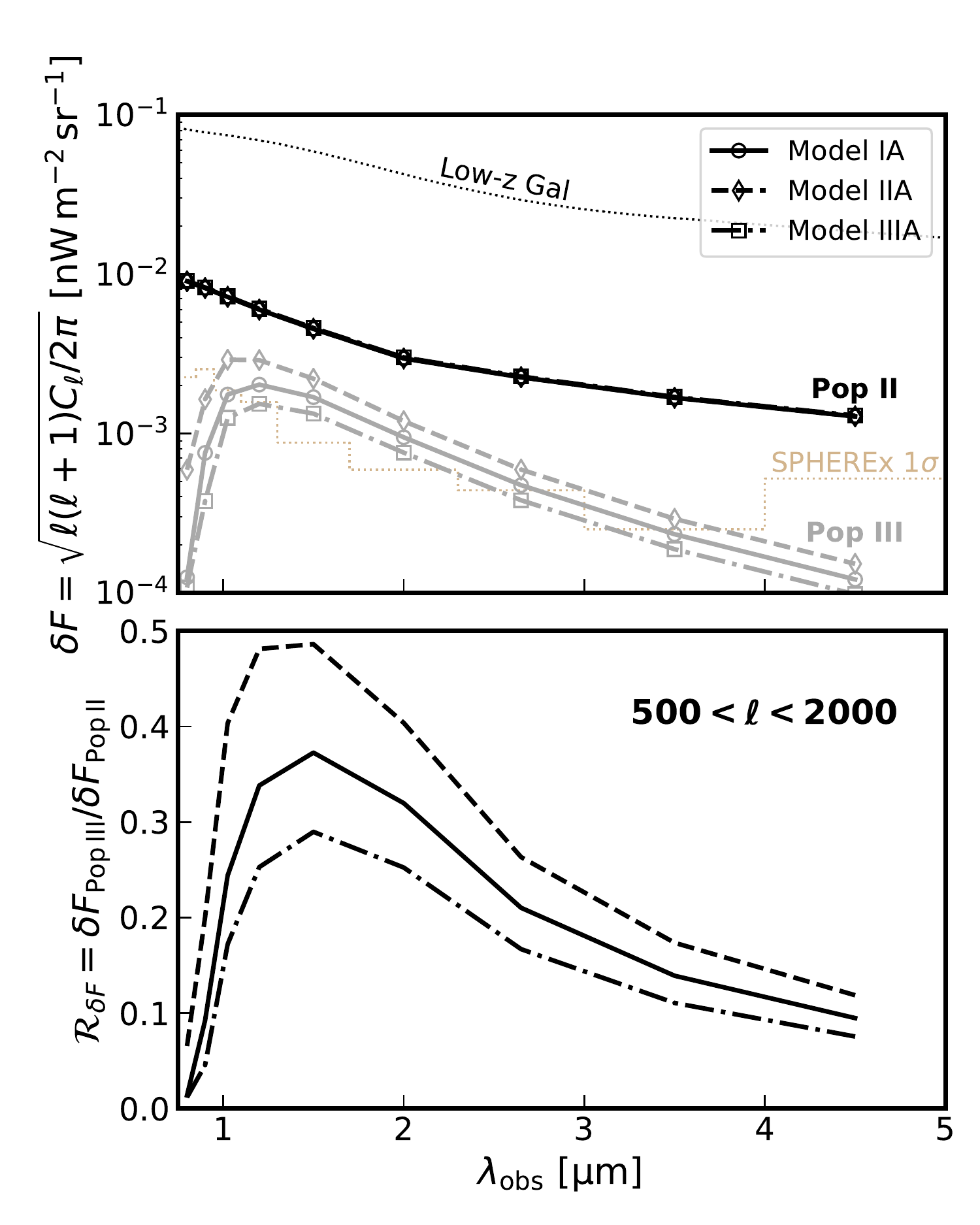}
 \caption{\textit{Top:} spectra of NIRB intensity fluctuations sourced by $z>5$ star-forming galaxies in the angular bin $500 < \ell < 2000$ predicted by the three variations of the Pop~II SFE $f_*$ defined in Table~\ref{tb:model_params}, compared with the broad-band uncertainties of the forthcoming survey in the $200\,\mathrm{deg^2}$ SPHEREx deep field. Also shown is the expected NIRB fluctuations contributed by low-$z$ galaxies after masking bright resolved sources, taken from \citet{Feng_2019}. \textit{Bottom:} the ratio of NIRB intensity fluctuations sourced by Pop~III and Pop~II stars. The strong evolution with wavelength is driven by the efficient production of Ly$\alpha$ emission by massive Pop~III stars.}
 \label{fig:dF_varypop2}
\end{figure}

To explore a range of plausible Pop~II SFHs, we consider two alternative ways of extrapolating the low-mass end of $f_*$ --- beyond the mass range probed by the observed UVLFs but still within the constraints of current data --- which are labeled as \texttt{steep} and \texttt{floor}, respectively, in Table~\ref{tb:model_params} following \citet{Mirocha_2017}.  

In Fig.~\ref{fig:dF_varypop2}, we show how the level of NIRB intensity fluctuations $\delta F$ and the Pop~III signature $\mathcal{R}_{\delta F}=\delta F_{\rm Pop~III}/\delta F_{\rm Pop~II}$ evolve with wavelength, as predicted by the three different combinations of our Pop~II SFE models and the reference Pop~III model, namely Model~IA, Model~IIA, and Model~IIIA. Values of $\delta F_{\rm Pop~II}$ and $\delta F_{\rm Pop~III}$ are quoted at the centres of the nine SPHEREx broadbands for multipoles $500 < \ell < 2000$ to facilitate a comparison with the $1\sigma$ surface brightness uncertainty of SPHEREx in each band, as illustrated by the staircase curve in tan (see Section~\ref{sec:results:sensitivity} for a detailed discussion of SPHEREx sensitivity forecasts). Overall, the imprint of Pop~III stars on the NIRB is connected to (and thus traces) their SFRD evolution through the strong Ly$\alpha$ emission they produced, with a peak/turnover at the wavelength of Ly$\alpha$ redshifted from the era when Pop~III star formation culminated/ended. Near the peak in the $1.5\,\mu$m band, the NIRB fluctuations contributed by Pop~III stars can be up to half as strong as the Pop~II contribution. Note that in practice the contribution of high-$z$ star-forming galaxies will be blended with other NIRB components from lower redshifts. Separation techniques relying on the distinction in the spectral shape of each component have been demonstrated in e.g., \citet{Feng_2019}. For reference, we show in Fig.~\ref{fig:dF_varypop2} the remaining fluctuation signal associated with low-$z$ ($z\lesssim3$) galaxies after masking bright, resolved ones, as predicted by the luminosity function model from \citet{Feng_2019}. Other sources of emission such as the IHL may also contribute a significant fraction of the total observed fluctuations --- though with a lower certainty, making the component separation even more challenging.

The effect of varying $f_*$ is pronounced for the Pop~III contribution, whereas the fluctuations sourced by Pop~II stars themselves are barely affected. As discussed in Section~\ref{sec:sfh_pop3} \cite[see also discussion in][]{Mebane_2018}, once formed in sufficient number, Pop~II stars can play an important role in shaping the Pop~III SFH by lifting the minimum mass of Pop~III haloes through their LW radiation. The contrast between the \texttt{steep} and \texttt{floor} models suggests that, for a fixed Pop~III model, changing $f_*$ within the range of uncertainty in UVLF measurements can vary the Pop~III signature on the NIRB by up to a factor of two. Unlike the Pop~III SFRD, whose dependence on $f_*$ grows over time as the LW background accumulates, the dependence of $\mathcal{R}_{\delta F
}$ on $f_*$ shows only modest evolution with wavelength since Pop~III stars formed close to the peak redshift dominate the fluctuation signal at all wavelengths. On the contrary, the Pop~II contribution remain almost unaffected by variations of $f_*$ because the majority of the fluctuation signal is contributed by Pop~II stars at $z\sim5$--6, which formed mostly in more massive haloes not sensitive to the low-mass end of $f_*$ (see Fig.~\ref{fig:SFRD}). 

\subsubsection{Effects of variations in the Pop~III SFH} \label{sec:results:imprints:pop3}

\begin{figure}
 \centering
 \includegraphics[width=0.48\textwidth]{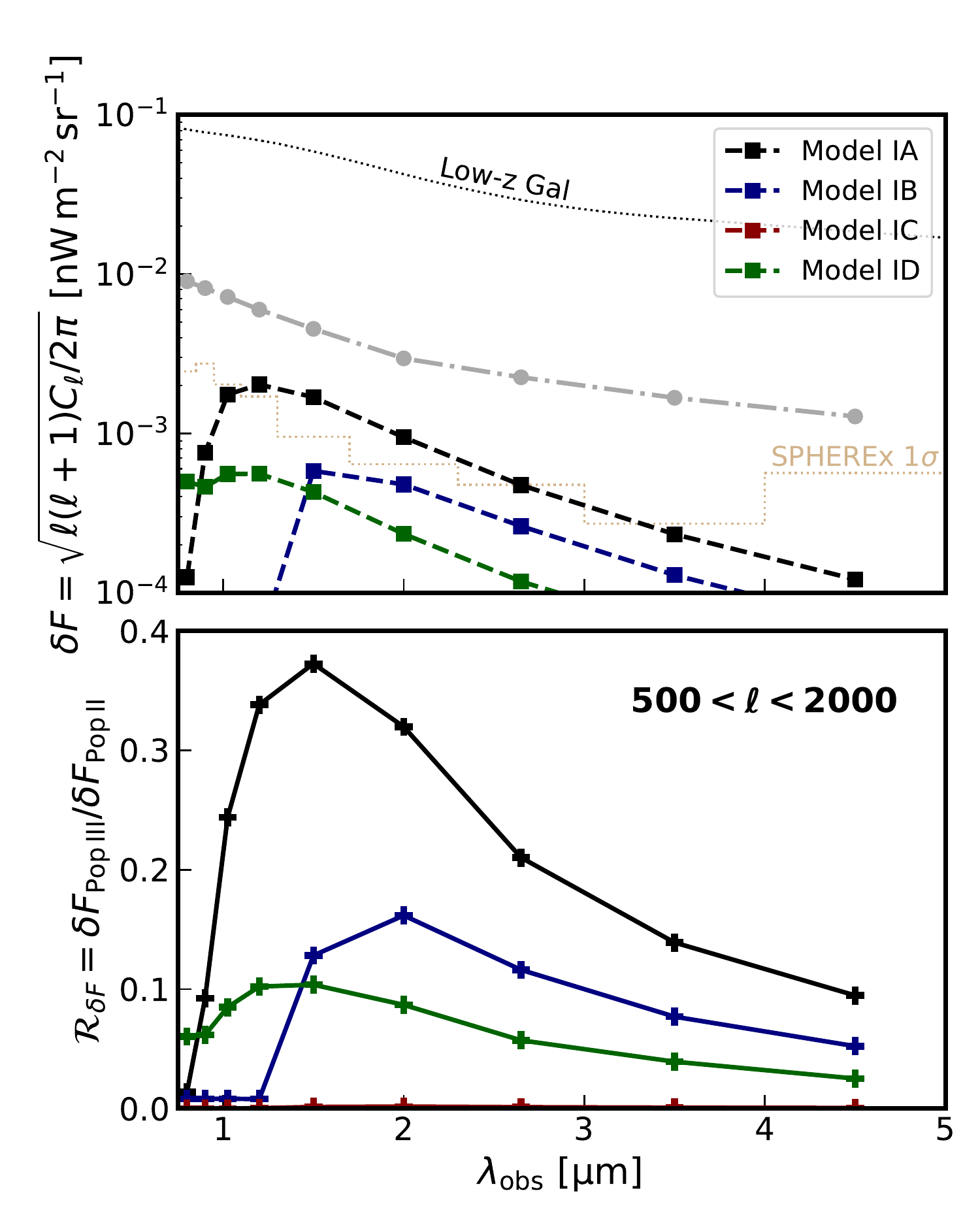}
 \caption{Same as Fig.~\ref{fig:dF_varypop2} but for the four variations of the Pop~III SFHs defined in Table~\ref{tb:model_params}. For comparison, the grey dash-dotted curve shows the Pop~II contribution to the fluctuations. }
 \label{fig:dF_varypop3}
\end{figure}

Apart from the influence of the LW background from Pop~II stars, the Pop~III SFH is also, and more importantly, determined by the physics of Pop~III star formation in minihaloes under the regulation of all sources of feedback. As specified in Table~\ref{tb:model_params}, we consider an additional set of three variations of the Pop~III star formation prescription and quantify how the imprint on the NIRB may be modulated. 

Similar to Fig.~\ref{fig:dF_varypop2}, Fig.~\ref{fig:dF_varypop3} shows the NIRB intensity fluctuations for the four different Pop~III models considered, each of which yields a possible Pop~III SFH fully regulated by the LW feedback and physical arguments about metal enrichment, as described in Section~\ref{sec:sfh_pop3}. Compared with the reference model (Model~IA), which implies an extremely high Pop~III star formation efficiency of order 0.1--1 by comparing rates of star formation and mass accretion, approximations to the semi-analytic models from \citet{Mebane_2018} imply less efficient Pop~III star formation and thus predict Pop~III SFRDs that are at least one order of magnitude smaller, as illustrated in Fig.~\ref{fig:SFRD}. Nevertheless, the fluctuation signals in Model~IB and ID are only a factor of 2--3 smaller than what Model~IA predicts, due to the high mass of Pop~III stars assumed in these models which yields a high photoionization rate of $Q(\mathrm{H})=10^{51}\,\mathrm{s}^{-1}$. Involving neither a high star formation efficiency ($\dot{M}^{\rm III}_*=3\times10^{-6}\,M_\odot\,\mathrm{yr}^{-1}$) nor a very top-heavy IMF ($Q(\mathrm{H})=10^{50}\,\mathrm{s}^{-1}$), Model~IC represents a much less extreme picture of Pop~III star formation favoured by some recent theoretical investigations \cite[e.g.,][]{Xu_2016,Mebane_2018}, which is unfortunately out of reach for any foreseeable NIRB measurement.

The correspondence between the Pop~III SFHs and their spectral signatures on the NIRB can be easily seen by comparing the shapes of $\dot{\rho}^\mathrm{III}_*(z)$ in Fig.~\ref{fig:SFRD} and $\mathcal{R}_{\delta F}$ in the bottom panel of Fig.~\ref{fig:dF_varypop3}, which suggests that the latter can be exploited as a useful probe for the efficiency and persistence of Pop~III formation across cosmic time. In particular, the detailed amplitude of $\mathcal{R}_{\delta F}$ is subject to astrophysical uncertainties associated with, e.g., the stellar SED and escape fraction, which are highly degenerate with the SFH as pointed out by \citetalias{FZ_2013}. However, the contrast between spectra showing turnovers at different redshifts (Model~IA vs Model~IB), or with or without a spectral break (Model~IB vs Model~ID), is robust, provided that the aforementioned astrophysical factors do not evolve abruptly with redshift. Any evidence for the existence of such a spectral signature from future facilities like SPHEREx would therefore be useful for mapping the landscape of Pop~III star formation. We further elaborate on the prospects for detecting the NIRB signal of Pop~III stars in the next subsection. 

\begin{table}
\centering
\caption{Survey and instrument parameters for SPHEREx deep field and CDIM medium field. Note that the surface brightness sensitivities are quoted at $1.5\,\mu$m for the $500 < \ell < 2000$ bin in the last row. The numbers inside the parentheses are the raw surface brightness sensitivities per $\ell$ mode per spectral resolution element, whereas the numbers outside are after spectral and spatial binning.}
\begin{tabular}{cccc}
\hline
Parameter & Description & SPHEREx & CDIM \\
\hline
$A_\mathrm{s}(\mathrm{deg}^2)$ & survey area & 200 & 30 \\
$R$ & resolving power & 40 & 300 \\
$f_{\rm sky}$ & sky coverage & 0.005 & 0.0007 \\
$\Omega_{\rm pix}(\mathrm{sr})$ & pixel size & $9.0\times10^{-10}$ & $2.4\times10^{-11}$ \\
$\sigma_{\rm pix}(\mathrm{nW/m^2/sr})$ & sensitivity (SB) & $0.09(1.94)$ & $0.14(24.06)$ \\
\hline
\end{tabular}
\label{tb:inst_params}
\end{table}

\subsection{Detecting Pop~III stars in the NIRB with SPHEREx and CDIM} \label{sec:results:sensitivity}

\begin{figure*}
 \centering
 \includegraphics[width=0.95\textwidth]{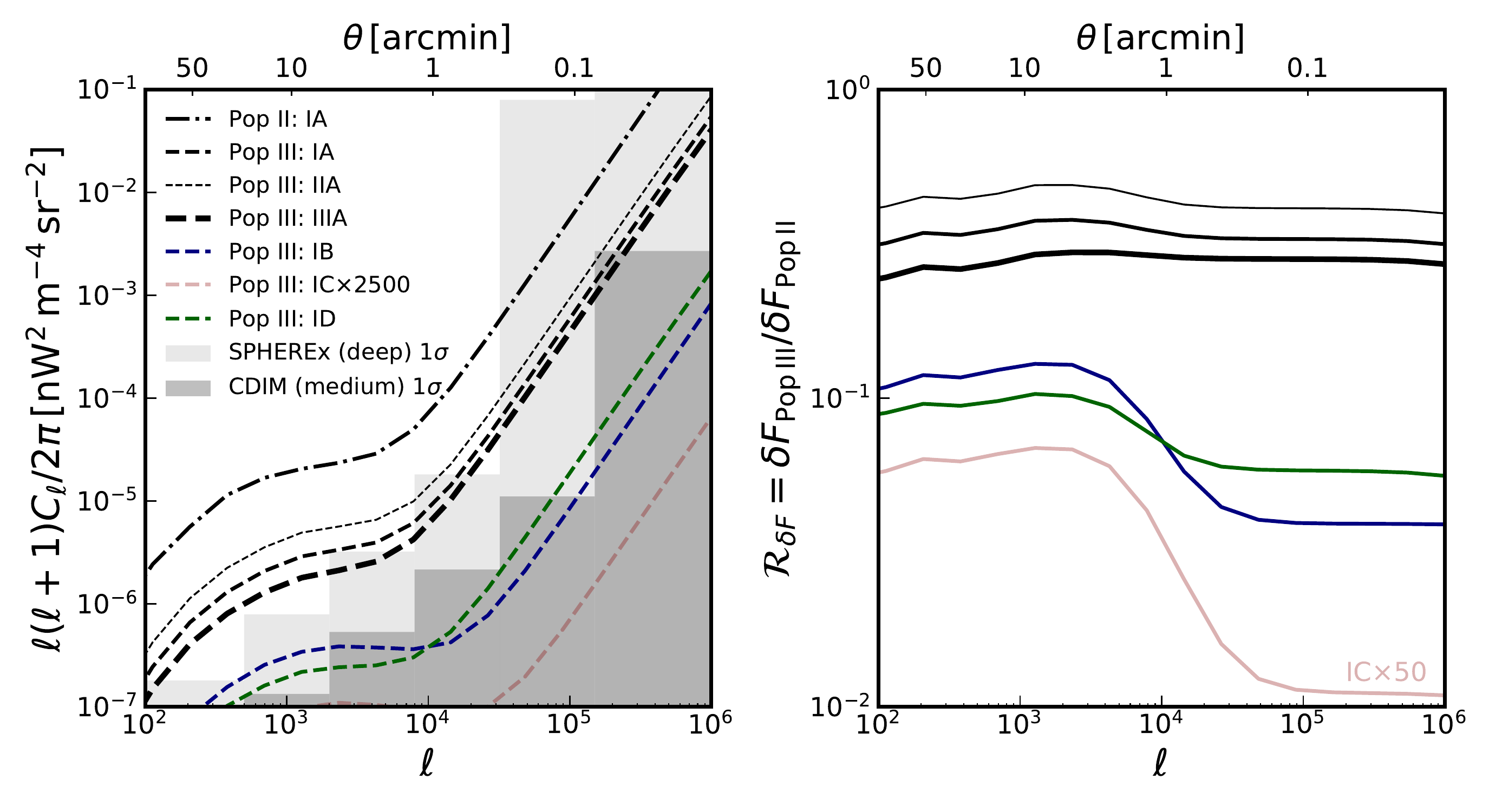}
 \caption{\textit{Left:} the angular auto-power spectrum $C_\ell$ of the NIRB at $1.5\,\mu$m predicted by different combinations of Pop~II and Pop~III models. Contributions from Pop~II and Pop~III stars are shown by dash-dotted and dashed curves, respectively. Variations of Pop~II model with \texttt{steep}, \texttt{dpl}, and \texttt{floor} SFE are represented by the thin, intermediate, and thick curves, respectively, whereas different colors represent different Pop~III variations. The prediction of Model~IC is raised by a factor of 2500 (50) to fit in the left (right) panel. The light and dark shaded regions indicate the expected band uncertainties of SPHEREx deep and CDIM medium surveys, respectively, after binning spectral channels and multipoles according to the imaging broadbands and angular bins defined (see text). Note that the band uncertainty of SPHEREx in the largest $\ell$ bin goes to infinity since such small scales are inaccessible, given the pixel size of SPHEREx. \textit{Right:} the ratio of NIRB intensity fluctuation amplitudes of Pop~III and Pop~II stars as a function of multipole moment $\ell$.}
 \label{fig:Cl_all}
\end{figure*}

To this point, we have elucidated how massive Pop~III stars might leave a discernible imprint on the observed NIRB when formed at a sufficiently high rate $\dot{M}^{\rm III}_{*} \gtrsim  10^{-3}\,M_\odot\,\mathrm{yr}^{-1}$ per minihalo whose minimum mass $M^{\rm III}_{h,\mathrm{min}}$ is set by the LW feedback, as well as how effects of varying Pop~II and Pop~III star formation physics can affect such a spectral signature. It is interesting to understand how well the NIRB signal contributed by high-$z$, star-forming galaxies may be measured in the foreseeable future, and more excitingly, what scenarios of Pop~III star formation may be probed. For this purpose, we consider two satellites that will be able to study the NIRB in detail, namely SPHEREx \citep{Dore_2014}, a NASA Medium-Class Explorer (MIDEX) mission scheduled to be launched in 2024, and CDIM \citep{Cooray_2019BAAS}, another NASA Probe-class mission concept. It is useful to point out that other experiments/platforms also promise to probe the NIRB signal from galaxies duration and before the EoR, including the ongoing sounding rocket experiment CIBER-2 \citep{Lanz_2014SPIE} and dedicated surveys proposed for other infrared telescopes such as JWST \citep{Kashlinsky_2015JWST} and Euclid \citep{Kashlinsky_2015Euclid}. In what follows, we focus on the forecasts for SPHEREx and CDIM given their more optimal configurations for NIRB observations, and refer interested readers to the papers listed for details of alternative methods. We note, though, that the high spectral resolution of CDIM (see Table~\ref{tb:inst_params}) makes 3D line-intensity mapping a likely more favourable strategy for probing first stars and galaxies than measuring $C_\ell$, when issues of foreground cleaning and component separation are considered. While in this work we only focus on the comparison of $C_\ell$ sensitivities, tomographic Ly$\alpha$ and H$\alpha$ observations with CDIM and their synergy with 21-cm surveys have been studied \cite[][]{Heneka_2017, HC_2021}.

Using the Knox formula \citep{Knox_1995}, we can write the uncertainty in the observed angular power spectrum $C_\ell$ measured for any two given bands as 
\begin{equation}
\Delta C_\ell = \frac{1}{\sqrt{f_{\rm sky}(\ell + 1/2)}} \left( C_\ell + C_\ell^\mathrm{noise} \right).
\label{eq:knox}
\end{equation}
The first term $C_\ell$ describes cosmic variance and the second term $C^{\rm noise}_\ell = 4\pi f_{\rm sky} \sigma^2_{\rm pix} N_{\rm pix}^{-1} e^{\Omega_{\rm pix}\ell^2}$ is the instrument noise \cite[][]{Cooray_2004}, where $N_{\rm pix}$ is the number of pixels in the survey. At sufficiently large scales where $\ell \ll \Omega^{-1/2}_{\rm pix}$, we have $C^{\rm noise}_\ell \approx \sigma^2_{\rm pix} \Omega_{\rm pix}$. The prefactor $[f_{\rm sky}(\ell + 1/2)]^{-1/2}$ accounts for the number of $\ell$ modes available, given a sky covering fraction of $f_{\rm sky}$. To estimate the instrument noise, we take the surface brightness sensitivity estimates made for a total survey area of $200\,\mathrm{deg}^2$ for SPHEREx and $30\,\mathrm{deg}^2$ for CDIM, corresponding to the deep- and medium-field surveys planned for SPHEREx and CDIM, respectively. The pixel size $\Omega_{\rm pix}$ is taken as $9.0\times10^{-10}\,\mathrm{sr}$ ($6.2\arcsec \times 6.2\arcsec$ pixels) and $2.4\times10^{-11}\,\mathrm{sr}$ ($1\arcsec \times 1\arcsec$ pixels) for SPHEREx and CDIM, respectively. 

Using the same spectral binning scheme as in \citet{Feng_2019}, we bin native spectral channels of both SPHEREx and CDIM into the following nine broadbands over an observed wavelength range of $0.75<\lambda_{\rm obs}<5\,\mu$m: $(0.75, 0.85)$, $(0.85, 0.95)$, $(0.95, 1.1)$, $(1.1, 1.3)$, $(1.3, 1.7)$, $(1.7, 2.3)$, $(2.3, 3.0)$, $(3.0, 4.0)$, and $(4.0, 5.0)$, regardless of their difference in the raw resolving power $R$ per channel. For the spatial binning of $\ell$ modes, we consider six angular bins over $10^2 < \ell < 10^6$ as follows: $(10^2, 5\times10^2)$, $(5\times10^2, 2\times10^3)$, $(2\times10^3, 8\times10^3)$, $(8\times10^3, 3\times10^5)$, $(3\times10^4, 1.5\times10^5)$,
and $(1.5\times10^5, 1\times10^6)$, which also apply to both SPHEREx and CDIM, although essentially no information is available on scales smaller than the pixel scale of the instrument. The $N=9$ broadbands specified then allow us to define an angular power spectrum vector $\boldsymbol{C}^{\bar{\lambda}_1\bar{\lambda}_2}_\ell$ (for each $\ell$ bin) that consists of $N(N+1)/2=45$ noise-included, auto- and cross-power spectra measurable from the broadband images. As shown in Table~\ref{tb:inst_params}, even though the surface brightness (SB) sensitivity per pixel of the CDIM medium field (T.-C.~Chang, private communication) is comparable to that of the SPHEREx deep field\footnote{See the public product for projected surface brightness sensitivity levels of SPHEREx at \url{https://github.com/SPHEREx/Public-products/blob/master/Surface_Brightness_v28_base_cbe.txt}.} after binning, its band noise power $C^{\rm noise}_\ell$ is in fact an order of magnitude lower thanks to CDIM's much smaller pixel size. For simplicity, we assume that the noise contribution from maps of different bands is uncorrelated, such that entries of the noise-included vector $\boldsymbol{\tilde{C}}^{\bar{\lambda}_1\bar{\lambda}_2}_\ell$ can be expressed as $C_\ell^{\bar{\lambda}_1\bar{\lambda}_2} + \delta_{\bar{\lambda}_1\bar{\lambda}_2}C_\ell^{\bar{\lambda}_1\bar{\lambda}_2,\mathrm{noise}}$, which is distinguished from the signal-only vector $\boldsymbol{C}^{\bar{\lambda}_1\bar{\lambda}_2}_\ell$ by the Kronecker delta $\delta_{\bar{\lambda}_1\bar{\lambda}_2}$. 

\begin{table}
\centering
\caption{The estimated raw S/N of NIRB signals sourced by Pop~II and Pop~III star-forming galaxies at $z>5$, using only auto-power spectra measured in the 9 broadbands or all 45 available auto- and cross-power spectra combined. For each entry, the first and second numbers represent the S/N estimated for SPHEREx deep survey and CDIM medium survey, respectively.}
\begin{tabular}{ccccc}
\hline
Model & $\mathrm{(S/N)}^{\rm auto}_{\rm Pop\,II}$ & $\mathrm{(S/N)}^{\rm auto}_{\rm Pop\,III}$ & $\mathrm{(S/N)}^{\rm all}_{\rm Pop~II}$ & $\mathrm{(S/N)}^{\rm all}_{\rm Pop~III}$  \\
\hline
IA & 68/1100 & 8.8/86 & 120/2300 & 13/110 \\
IB & 68/1100 & 1.9/38 & 120/2300 & 2.8/45 \\
IC & 68/1100 & 0.0/$1\times10^{-3}$ & 120/2300 & 0.0/$2\times10^{-3}$ \\
ID & 68/1100 & 0.8/6.0 & 120/2300 & 1.4/10 \\
\hline
\end{tabular}
\label{tb:snr}
\end{table}

The resulting signal-to-noise ratio (S/N) of the full-covariance measurement (summed over all angular bins of $\ell$)
\begin{equation}
\left( \frac{S}{N} \right)^2 = \sum_{\ell} \left( \boldsymbol{C}^{\bar{\lambda}_1\bar{\lambda}_2}_\ell \right)^{\rm T} \left( \boldsymbol{C}^{\bar{\lambda}_1 \bar{\lambda}_2, \bar{\lambda}_1^\prime \bar{\lambda}_2^\prime}_{\ell,\mathrm{COV}} \right)^{-1} \left( \boldsymbol{C}^{\bar{\lambda}_1^\prime\bar{\lambda}_2^\prime}_\ell \right)~
\label{eq:C_SNR}
\end{equation}
is then used to quantify the detectability of the NIRB signals by the two surveys considered. Here, the covariance matrix between two band power spectra $\boldsymbol{C}^{\bar{\lambda}_1\bar{\lambda}_2}_\ell$ and $\boldsymbol{C}^{\bar{\lambda}_1^\prime\bar{\lambda}_2^\prime}_\ell$ can be expressed using Wick's theorem as \cite[][]{Feng_2019}
\begin{equation}
\boldsymbol{C}^{\bar{\lambda}_1 \bar{\lambda}_2, \bar{\lambda}_1^\prime \bar{\lambda}_2^\prime}_{\ell,\mathrm{COV}} = \frac{1}{f_{\rm sky}(2\ell + 1)} \left[ \tilde{C}^{\bar{\lambda}_1\bar{\lambda}^\prime_1}_\ell \tilde{C}^{\bar{\lambda}_2\bar{\lambda}^\prime_2}_\ell + \tilde{C}^{\bar{\lambda}_1\bar{\lambda}^\prime_2}_\ell \tilde{C}^{\bar{\lambda}_1^\prime\bar{\lambda}_2}_\ell\right],
\end{equation}
which reduces to equation~(\ref{eq:knox}) when $\bar{\lambda}_1 = \bar{\lambda}_1^\prime = \bar{\lambda}_2 = \bar{\lambda}_2^\prime$. 

\begin{figure*}
  \begin{minipage}[b]{0.5\linewidth}
    \centering
    \includegraphics[width=.95\linewidth]{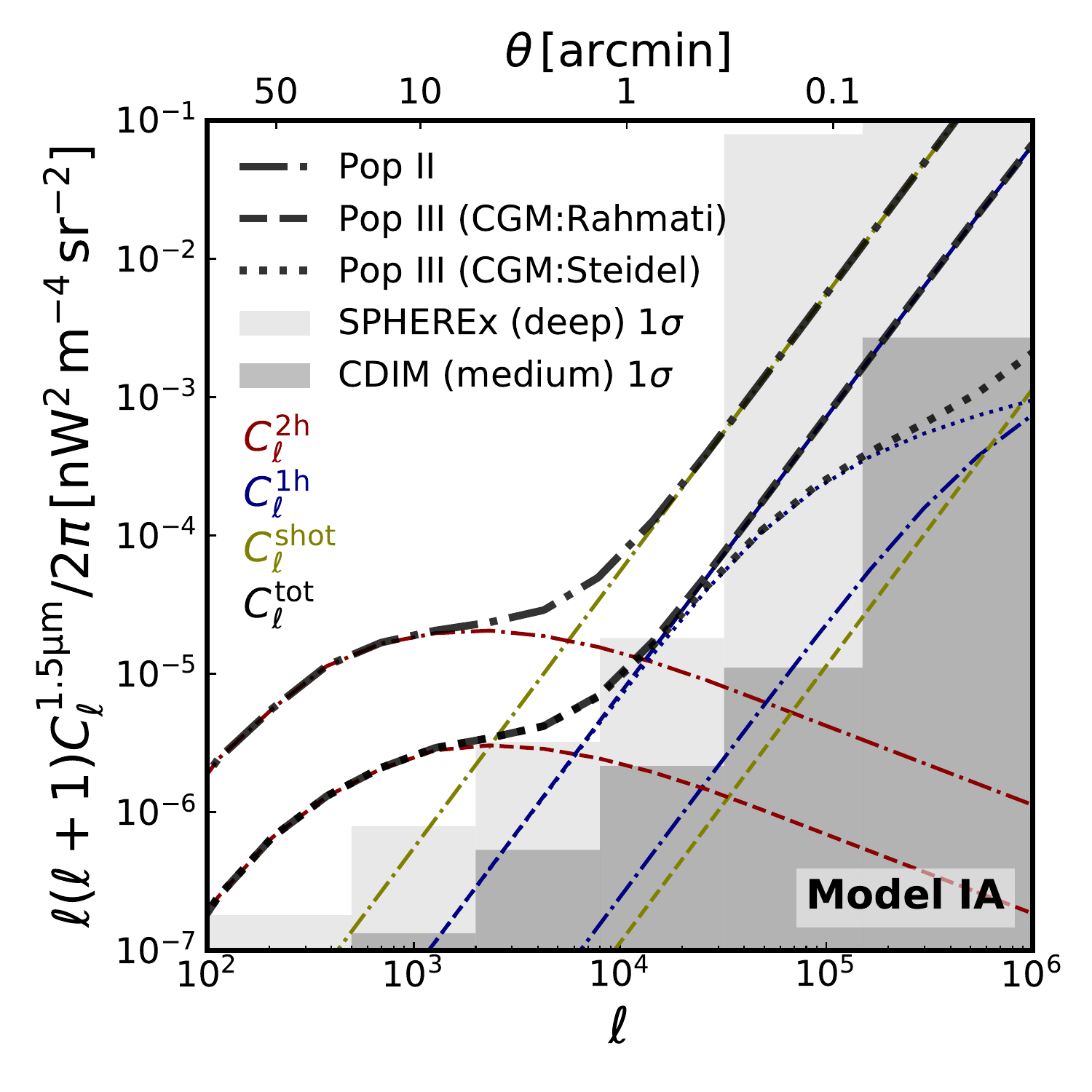} 
    \vspace{1ex}
  \end{minipage}%%
  \begin{minipage}[b]{0.5\linewidth}
    \centering
    \includegraphics[width=.95\linewidth]{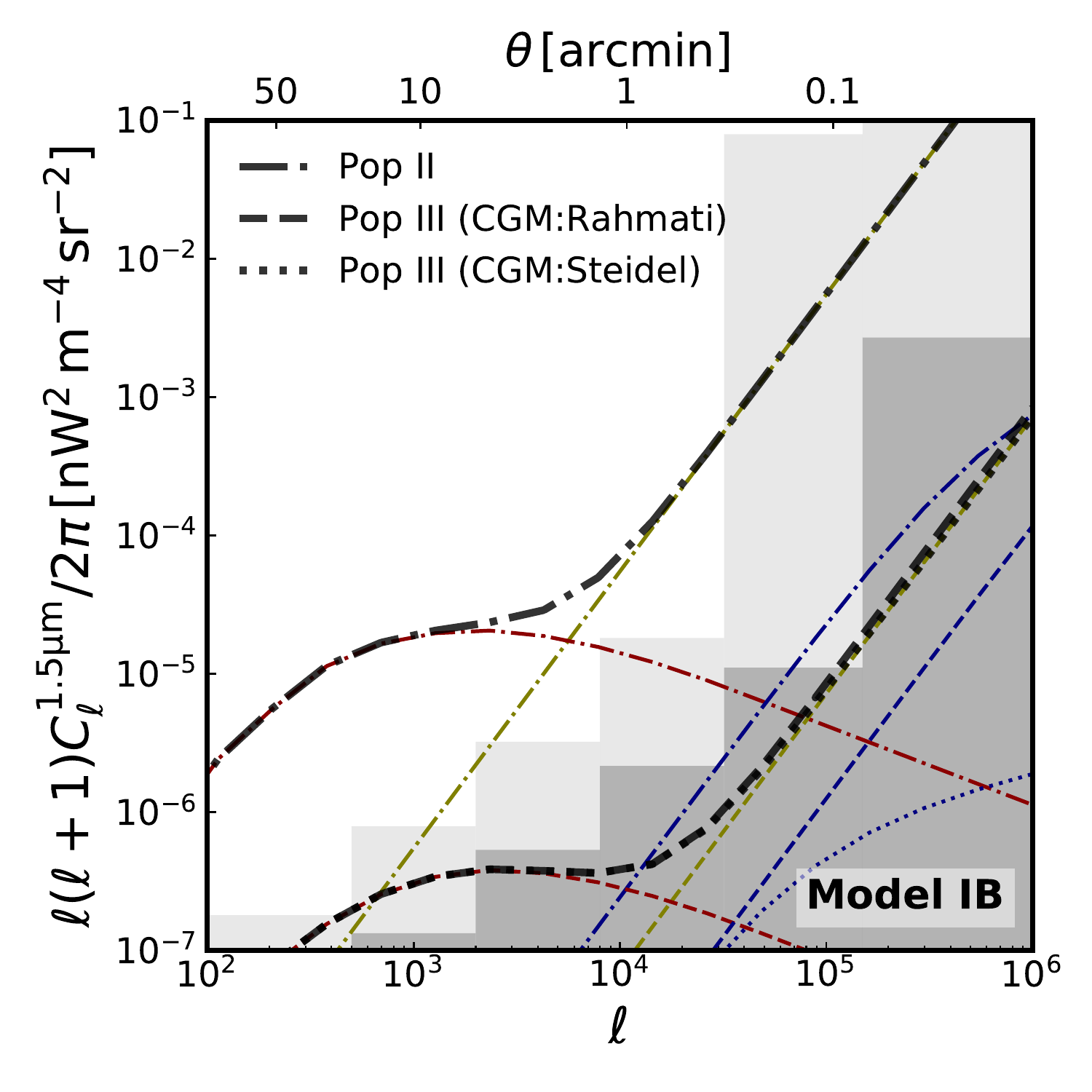}
    \vspace{1ex}
  \end{minipage} 
  \begin{minipage}[b]{0.5\linewidth}
    \centering
    \includegraphics[width=.95\linewidth]{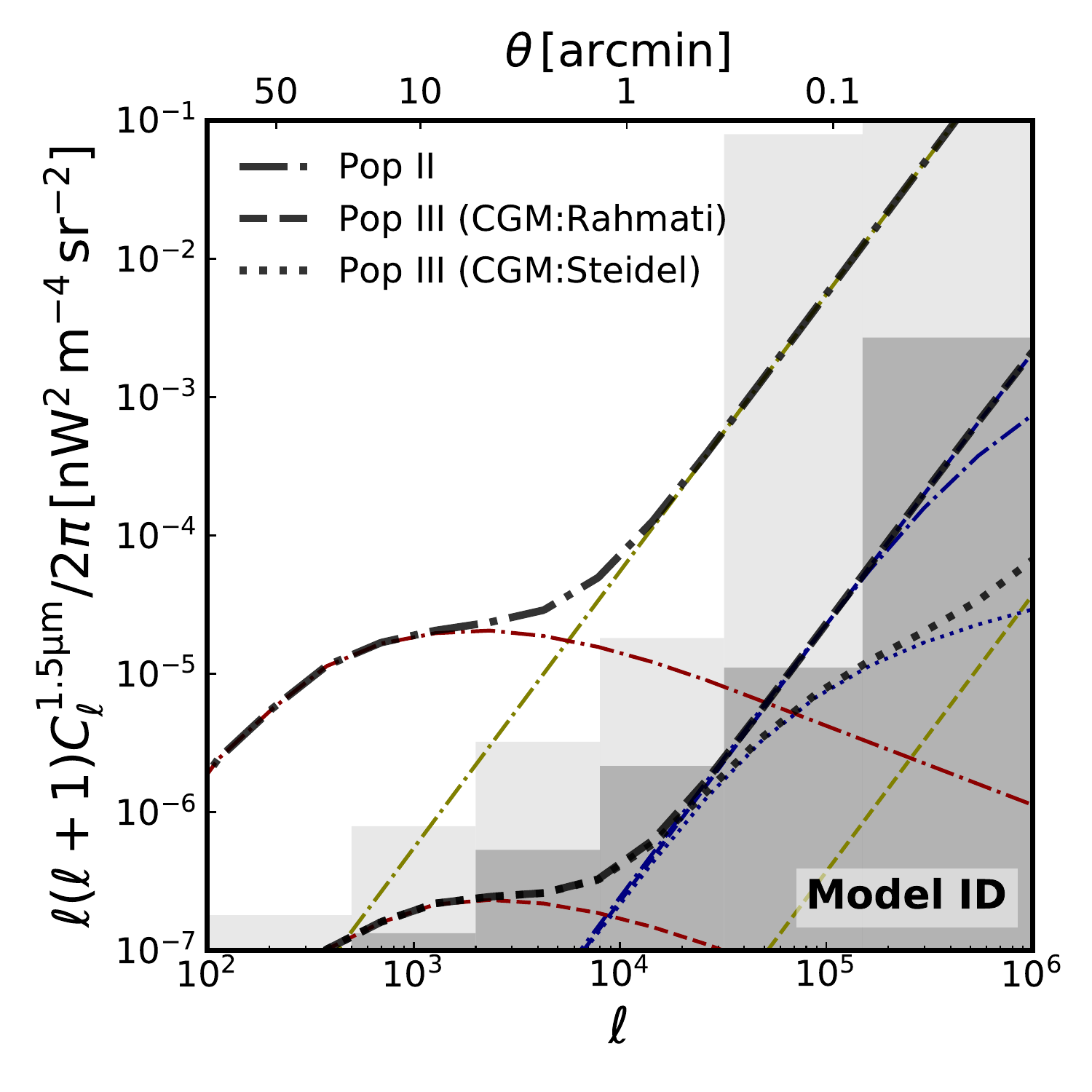} 
    \vspace{1ex}
  \end{minipage}%% 
  \begin{minipage}[b]{0.5\linewidth}
    \centering
    \includegraphics[width=.95\linewidth]{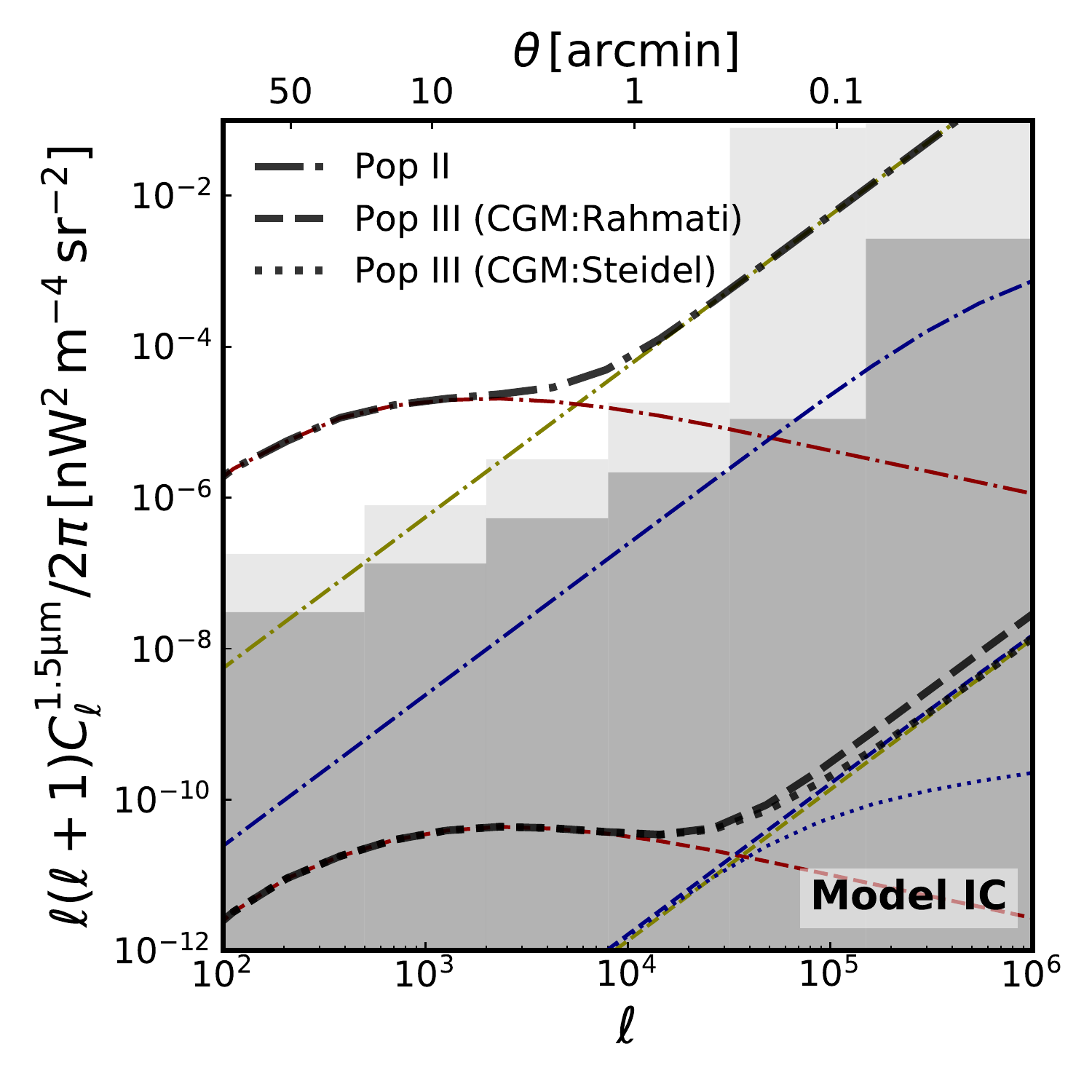} 
    \vspace{1ex}
  \end{minipage} 
 \caption{Contributions from $z>5$ Pop~II and Pop~III star-forming galaxies to the two-halo, one-halo and shot-noise components of $C_\ell$ measured at 1.5\,$\mu$m. Clockwise from the top left panel: the figures show $C_\ell$ predicted by Model~IA, Model~IB, Model~IC, and Model~ID, defined in Table~\ref{tb:model_params}. In each panel, the one-halo term is shown for two instances of CGM profile to illustrate the connection between the escape of ionizing photons and the shape of the one-halo term. The light and dark shaded regions indicate the expected band uncertainties of SPHEREx deep and CDIM medium surveys, respectively, after binning spectral channels and multipoles according to the imaging broadbands and angular bins defined (see text). Note the different $y$-axis scale used in the bottom right panel to show the Pop~III signal.}
 \label{fig:Cl} 
\end{figure*}

In Table~\ref{tb:snr}, we summarize the raw sensitivities to $C_\ell$ in terms of the total S/N that SPHEREx and CDIM are expected to achieve in the four different Pop~III star models considered in this work. Since the contribution from Pop~II stars dominates over that from Pop~III stars at all wavelengths except where the Pop~III signature appears ($\sim 1.5\,\mu$m), a significantly higher raw S/N is expected for the former, reaching above 100 when combining all the auto- and cross-correlations available and summing up all angular bins for SPHEREx, similar to what was previously found by \citet{Feng_2019}. For Pop~III stars, our optimistic Model~IA predicts a raw S/N greater than 10 for SPHEREx, which is dominated by the first three angular bins with $\ell \lesssim 10^4$, whereas more conservative models assuming lower Pop~III SFR per halo predict much smaller raw S/N of only a few. Compared with SPHEREx, CDIM is expected to provide 
approximately a factor of 20 (10) improvement on the total (Pop~III) raw S/N achievable, thanks to the competitive SB sensitivity at its small pixel size. This allows CDIM to measure the Pop~III contribution at the same significance (S/N $\sim100$) as the Pop~II contribution for SPHEREx in Model~IA when the full covariance is leveraged. We note, though, that in practice the contribution from high-$z$, star-forming galaxies must be appropriately separated from all other components of the source-subtracted NIRB, such as unsolved low-$z$ galaxies, the IHL and the diffuse Galactic light (DGL), which lead to a significant reduction of the constraining power on the high-$z$ component \citep{Feng_2019}. This component separation issue will be discussed further in Section~\ref{sec:discussion:separation}. 

We show in the left panel of Fig.~\ref{fig:Cl_all} a comparison of the auto-correlation angular power spectra $C_\ell$ of the NIRB predicted by our models in the $1.5\,\mu$m band. For clarity, we only show the Pop~II signal in Model~IA (dash-dotted curve) since it hardly varies with the model variations considered. For Pop~III stars, a subset of models yielding NIRB signals potentially detectable for SPHEREx and/or CDIM are shown by the dashed curves with varying thickness and color. The pessimistic Model~IC is rescaled and then plotted for completeness. The right panel of Fig.~\ref{fig:Cl_all} illustrates the effect of changing Pop~II and Pop~III models on the shape of $C_\ell$ by showing the ratio of intensity fluctuations $\mathcal{R}_{\delta F}$, which is used to characterize the Pop~III signature in Section~\ref{sec:results:imprints}, as a function of $\ell$. In all models, $\mathcal{R}_{\delta F}$ peaks at around $\ell \sim 10^3$ or an angular scale of $\sim10$\arcmin, similar to what was found by e.g., \citet{Cooray_2004}. The fact that in cases like Model~IB the fluctuations are preferentially stronger on large angular scales compared to Model~IA is because, in the former case, Pop~III stars formation completed at much higher redshift and thus was more clustered. 

In Fig.~\ref{fig:Cl}, we further show the halo-model compositions (i.e., one-halo, two-halo and shot-noise terms) of $C_\ell$ in each Pop~III model. Moreover, two possible forms of the one-halo profile motivated by the CGM models, as described in Section~\ref{sec:nirb_fluctuation} and Fig.~\ref{fig:fprofile}, are displayed for the Pop~III contribution. 

\begin{figure*}
 \centering
 \includegraphics[width=0.99\textwidth]{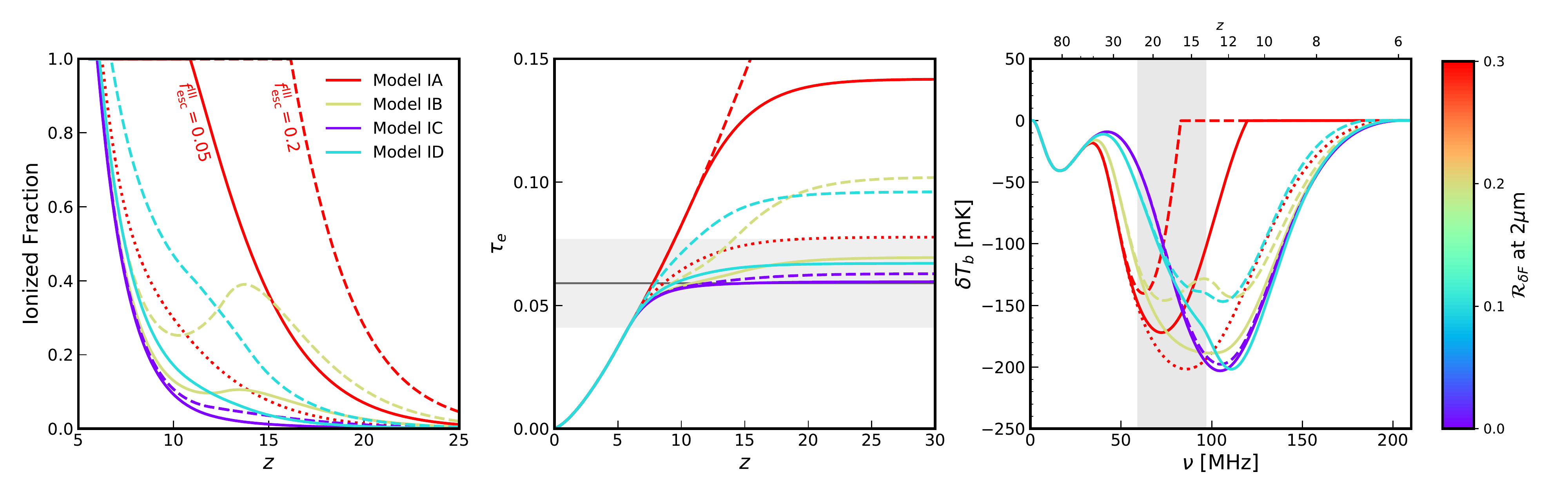}
 \caption{\textit{Left:} impact of Pop~III stars on the reionization history and NIRB fluctuations. Different line styles represent different assumptions of $f^{\rm III}_{\rm esc}$, with an additional dotted curve showing the case of $f^{\rm III}_{\rm esc}=0.01$ for Model~IA. Curves are color-coded by the Pop~III signature $\mathcal{R}_{\delta F}$ at $2\,\mu$m, where the models can be best distinguished from each other. \textit{Middle:} the electron scattering optical depth $\tau_e$ implied by each model. The horizontal line and grey shaded region indicate the $3\sigma$ confidence interval on $\tau_e$ inferred from CMB polarization data measured by Planck \citep{Pagano_2020}. \textit{Right:} the 21-cm global signal $\delta T_b$ implied by each model. The grey shaded region indicates the width of the global signal peaking at 78\,MHz as measured by EDGES \citep{Bowman_2018}.}
 \label{fig:implications}
\end{figure*}

Three notable features show up from this decomposition of $C_\ell$. First, the relative strengths of the one-halo component $C^\mathrm{1h}_\ell$ and shot-noise component $C^\mathrm{shot}_\ell$ are distinct for Pop~II and Pop~III stars. Because the nebular emission is subdominant to the stellar emission for Pop~II stars, on small angular scales their one-halo term is negligible compared to the shot-noise term, making $C_\ell$ of Pop~II stars almost scale-invariant at $\ell > 10^4$. On the contrary, Pop~III stars can produce very strong nebular emission, especially Ly$\alpha$, which makes it possible for their one-halo term to dominate on small angular scales. Such an effect can be seen in the left two panels of Fig.~\ref{fig:Cl}, where the one-halo term is approximately 1.5\,dex higher than the shot-noise term. 

Second, amplitudes of the one-halo and shot-noise components also depend on the exact SFH, or more specifically, the persistence of Pop~III star formation. As shown by the contrast between the left and right two panels of Fig.~\ref{fig:Cl}, models with an extended Pop~III SFH (but not necessarily a later Pop~III to Pop~II transition, see Fig.~\ref{fig:SFRD}) that persists till $z<10$ provide the nebular emission with sufficient time to overtake the stellar emission in the contribution to the NIRB, thereby resulting in a stronger one-halo term. 

Last but not least, we leverage the physical picture illustrated in Fig.~\ref{fig:fprofile} to enable additional flexibility in the modelling of the one-halo term by physically connecting its profile with the escape fraction of ionizing photons $f^{\rm III}_{\rm esc}$. Taking the two CGM models considered and described in Section~\ref{sec:nirb_fluctuation}, we get two distinct profiles corresponding to (lower limits on) escape fractions of 5\% and 20\%, respectively. When the one-halo term is strong enough on scales of $\ell > 10^4$, e.g., in Model~IA or ID, such a difference in the radiation profile leads to a clear distinction in the shape of the total power spectrum on these scales. This can be seen by comparing the dashed and dotted curves in black in Fig.~\ref{fig:Cl}, with a more scale-dependent one-halo term corresponding to a more extended profile of ionizing flux and thus higher escape fraction. It is useful to note that, in most cases considered in this work, an escape fraction of 20\% for Pop~III stars ends up with a reionization history too early to be consistent with the CMB optical depth constraint from the Planck polarization data, as we will discuss in the next section. Nevertheless, we consider that the two values of $f^{\rm III}_{\rm esc}$ chosen are plausible, allowing us to demonstrate how constraints on small-scale fluctuations, in particular the detailed shape of $C^\mathrm{1h}_\ell$, that SPHEREx and CDIM are likely to place may shed light on the escape of ionizing photons from the first ionizing sources at $z \gtrsim 10$. 

To this point, we have shown how detectable the high-$z$ contribution from Pop~II and Pop~III stars to the NIRB would be when compared with sensitivity levels achievable by upcoming/proposed instruments. An important question that follows is how to separate this high-$z$ component from others and, preferably, disentangle the Pop~II and Pop~III signals. Without the input of external data sets, such as another tracer of star-forming galaxies to be cross-correlated with, the key idea of the solution lies in the utilization of the distinctive spatial and spectral structures of different components. As shown in Fig.~\ref{fig:nuInu}, the high-$z$ component dominated by Pop~II stars is characterized by a Lyman break due to the blanketing effect of intergalactic \ion{H}{i}. Such a spectral feature has been demonstrated to be useful for isolating the high-$z$ component from sources from lower redshifts \cite[e.g.,][]{Feng_2019}. Similar ideas apply to the separation of the much weaker Pop~III signal from the Pop~II signal, thanks to distinctions in their wavelength dependence (due to different types of emission dominating Pop~II and Pop~III signals, see Fig.~\ref{fig:dF_varypop3}) and angular clustering (due to different halo mass and redshift distributions of Pop~II and Pop~III signals, see Fig.~\ref{fig:Cl_all}). Despite an extremely challenging measurement, these contrasts in spatial and spectral structures make it possible, at least in principle, to distinguish templates of the high-$z$ component as a whole or Pop~II and Pop~III signals separately. We will elaborate on this component separation issue further in Section~\ref{sec:discussion:separation}, although a detailed study of it is beyond the scope of this paper and thus reserved for future work. 

\section{Implications for Other Observables} \label{sec:imply}

Probing ionizing sources driving the EoR with an integral and statistical constraint like the NIRB has a number of advantages compared to the observation of individual sources, including lower cost of observing time, better coverage of the source population, and importantly, synergy with other observables of the EoR. Taking our models of high-$z$ source populations for the NIRB, we discuss in this section possible implications for other observables, such as the reionization history and 21-cm signal, that can be made from forthcoming NIRB measurements. 

\subsection{Reionization history}
\label{sec:imply-reionization}

In the left panel of Fig.~\ref{fig:implications}, we show reionization histories, characterized by the volume-averaged ionized fraction of the IGM, that our models of Pop~II/III star formation predict under two different assumptions of the escape fraction $f^{\rm III}_{\rm esc}$, namely 5\% and 20\% derived from the CGM models by \citet{Rahmati_2015} and \citet{Steidel_2010}, respectively. We note that to compute the reionization history, we assume a constant escape fraction of $f^{\rm II}_{\rm esc}=10\%$ for Pop~II stars, which is known to yield a $\tau_e$ in excellent agreement with the best-estimated value based on the latest Planck data \cite[e.g.,][]{Pagano_2020} without Pop~III contribution. The middle panel of Fig.~\ref{fig:implications} shows contributions to the total $\tau_e$ at different redshifts calculated from the reionization histories predicted. Among the four models shown, Model~IA forms Pop~III stars too efficiently to reproduce the $\tau_e$ constraint from Planck, even with $f^{\rm III}_{\rm esc}$ as low as 5\%. To reconcile this tension, we include an additional case setting $f^{\rm III}_{\rm esc}$ to 1\% as shown by the red dotted curve, which yields a $\tau_e$ value marginally consistent with the Planck result. We stress that the LyC escape fraction of Pop~III galaxies is poorly understood. A ``radiation-bounded'' picture of the escape mechanism generally expects an higher escape fraction than Pop~II galaxies, due to the extremely disruptive feedback of Pop~III stars \citep{Xu_2016ESC}. A ``density-bounded'' picture, however, requires the ionized bubble to expand beyond the virial radius, and thus predicts significantly lower LyC escape fraction for relatively massive ($M_h \gtrsim 10^{6.5}\,M_\odot$) minihaloes where the majority of Pop~III stars formed \cite[e.g.,][]{TH_2021}. Therefore, besides $\tau_e$, which is arguably the most trusted observable, NIRB observations provide an extra handle on jointly probing the SFR and escape fraction of minihaloes forming Pop~III stars.

In general, earlier reionization is expected for a model that predicts stronger Pop~III star signature on the NIRB, and in Model~IB, where the Pop~III to Pop~II transition is early and rapid, unusual double reionization scenarios can even occur. A caveat to keep in mind, though, is that certain forms of feedback, especially photoheating, that are missing from our model can actually alter the chance of double reionization by affecting the mode and amount of star formation in small haloes, making double reionization 
implausible \cite[][]{Furlanetto_2005}. As such, we refrain from reading too much into this double reionization feature, which is likely due to the incompleteness of our modelling framework, and focus on the integral measure $\tau_e$ instead. While it is challenging to establish an exact mapping between the NIRB signal and reionization history, detecting a Pop~III signal as strong as what Model~IB or ID predicts would already provide tantalizing evidence for a nontrivial contribution to the progression of reionization from Pop~III stars. Such a high-$z$ tail for reionization may be further studied through more precise and detailed measurements of NIRB imprints left by Pop~III stars, or via some alternative and likely complementary means such as the kSZ effect \cite[e.g.,][]{Alvarez_2021} and the E-mode polarization of CMB photons \cite[e.g.,][]{Qin_2020, Wu_2021}. Also worth noting is that, in order not to overproduce $\tau_e$, in cases where the Pop~III signature is nontrivial the escape fraction must be either restricted to a sufficiently small upper bound, or allowed to evolve with halo mass and/or redshift. Such constraints on the form of $f^{\rm III}_{\rm esc}$ would become more stringent for a stronger NIRB signature, as indicated by the curves in different colors and line styles in the middle panel of Fig.~\ref{fig:implications}. Combining measurements of $C^\mathrm{1h}_\ell$ on sub-arcmin scales with observations of the EoR history, we find it possible to constrain the budget of ionizing photons from Pop~III stars, especially $f^{\rm III}_{\rm esc}$. 

\subsection{The 21-cm signal}

We show in the right panel of Fig.~\ref{fig:implications} the 21-cm global signal, i.e., the sky-averaged differential brightness temperature of the 21-cm line of neutral hydrogen, implied by each of our Pop~III star formation models. Similar to what is found by \citet{Mirocha_2018}, models with efficient formation of massive Pop~III stars, which leave discernible imprints on the NIRB, predict qualitatively different 21-cm global signals from that predicted by a baseline model without significant Pop~III formation (e.g., Model~IC). Except for cases with unrealistically early reionization, Pop~III stars affect the low-frequency side of the global signal the most, modifying it into a broadened and asymmetric shape that has a high-frequency tail. The absorption trough gets shallower with increasing Pop~III SFR and/or $f^{\rm III}_{\rm esc}$, as a result of enhanced heating by the X-rays and a lower neutral fraction. 

A tentative detection\footnote{Note, however, that concerns remain about the impact of residual systematics such as foreground contamination on the EDGES results \cite[see e.g.,][]{Hills_2018, DM_2018, Bradley_2019, SP_2020}.} of the 21-cm global signal was recently reported by the Experiment to Detect the Global Epoch of Reionization Signature (EDGES; \citealt{Bowman_2018}), which suggests an absorption trough centered at 78.1\,MHz, with a width of 18.7\,MHz and a depth of more than $-500$\,mK. Regardless of the absorption depth, which may only be explained by invoking some new cooling channels of the IGM or some additional radio sources (than the CMB) in the early universe, a peak centering at 78.1\,MHz is beyond the expectation of simple Pop~II-only models based on extrapolations of the observed galaxy UVLFs \citep{MF_2019}. Additional astrophysical sources such as Pop~III stars may help provide the early Wouthuysen–Field (WF) coupling effect and X-ray heating required to explain the absorption at 78.1\,MHz, as shown in the right panel of Fig.~\ref{fig:implications} by the shift of curves towards lower frequencies \cite[see also][]{Mebane_2020}. Therefore, insights into the Pop~III SFH from NIRB observations would be highly valuable for gauging how much the tension between the EDGES signal and galaxy model predictions might be reconciled by including the contribution of Pop~III stars. 

Besides the global signal, fluctuations of the 21-cm signal also serves as an important probe of reionization. Various physical properties of Pop~III stars are expected to be revealed through their effects on cosmic 21-cm power spectrum, especially the timings of the three peaks corresponding to WF coupling, X-ray heating, and reionization (Mebane et al. in prep). On the other hand, the cross-correlation between 21-cm and NIRB observations has been discussed in a few previous works as a way to trace the reionization history \cite[e.g.,][]{Fernandez_2014, Mao_2014}. We will investigate how to develop a much deeper understanding of Pop~III star formation from synergies of 21-cm and NIRB data in future work.

% -------------------------- S5: Discussion -------------------------- %

\section{Discussion} \label{sec:discussion}

\subsection{Limitations and the sensitivity to model assumptions} \label{sec:discussion:limitation}

So far, we have described a semi-empirical model of the high-$z$ NIRB signal, based on physical arguments of Pop~II and Pop~III star formation calibrated against latest observations of high-$z$ galaxies. Our modelling framework, however, is ultimately still simple in many ways. While more detailed treatments are beyond the scope of this paper and thus left for future work, in what follows, we discuss some major limitations of our model, together with how our findings might be affected by the simplified assumptions. 

A key limitation of our model is its relatively simple treatment of the emission spectra of source populations. Despite that (i) the Pop~II SED is modelled with the SPS, assuming the simplest possible composite stellar population with a constant SFH, and (ii) the Pop~III SED can be reasonably approximated as a blackbody, certain aspects of the complicated problem are unaccounted. These include choices of the IMF, stellar metallicity (for Pop~II stars only) and age, etc. and their potential redshift evolution, as well as effects of the stochasticity among galaxies, the extinction by dust, and so forth. We expect our main results about Pop~III stars, phrased in terms of a ``perturbation'' to the Pop~II-only baseline scenario, to be robust against these sources of complexity, even though quantifying their exact effects on the shape and amplitude of high-$z$ NIRB signals would be highly valuable in the near future. 

Another important limitation is associated with free parameters that are loosely connected to the physics of source populations, such as the nuisance parameters defining the shape of $f_*$, escape fractions of LyC and LW photons, and parameters $\mathcal{T}_c$ and $\mathcal{E}_c$ used to set the efficiency and persistence of Pop~III star formation. While making it easy to explore a wide range of possible scenarios of star formation and reionization, these parameters may not represent an ideal way to parameterize the high-$z$ NIRB signal, meaning that they can be oversimplified or physically related to each other and other implicit model assumptions such as the IMF in practice. Either way, unwanted systematics and degeneracy could arise, making data interpretation with the model challenging and less reliable. Looking ahead, we find it useful to develop a more unified (but still flexible) framework for parameterizing the NIRB, identifying and reflecting the connections among physical quantities/processes of interest. This will be particularly useful for parameter inference in the future. 

\subsection{Component separation of the observed NIRB} \label{sec:discussion:separation}

As already mentioned at the end of Section~\ref{sec:results:sensitivity}, an important challenge in the NIRB data analysis is the separation of its components, which have a broad range of astrophysical origins \cite[][]{Kashlinsky_2018RvMP}. Failing to perform component separation properly and effectively will make it impossible to constrain a component as weak as the signal from high-$z$ galaxies. Fortunately, as demonstrated in \citet{Feng_2019}, by measuring the full-covariance angular power spectrum of the observed NIRB, one can reliably separate the major components thanks to their different spatial and spectral structures. In the presence of much stronger low-$z$ components, this approach allows the contribution from EoR galaxies to be recovered and constrained with sufficient significance (S/N$\ \gtrsim 5$), without the need for external data sets. To actually reveal the formation histories of the first stars, one must also tell apart the contributions of Pop~II and Pop~III stars. In addition to the similar full-covariance method discussed in Section~\ref{sec:results:sensitivity}, which makes use of the spectral and spatial differences of Pop~II and Pop~III signals, it can be also promising to consider a joint analysis with ancillary data. External datasets such as 21-cm maps (e.g., Cox et al., in prep) and galaxy distributions \cite[e.g.,][]{Scott_2021arXiv} can be useful resources for cross-correlation analyses, which are expected to be available from observatories such as HERA \citep{DeBoer_2017HERA}, SKA \citep{Mellema_2013SKA}, and the Roman Space Telescope \citep{Spergel_2015arXiv} in the coming decade. While tracers like the 21-cm signal and photometric galaxies are also complicated by foregrounds and/or survey-specific systematics, which cause loss of information in inaccessible modes, the extra redshift information from cross-correlating the NIRB with these 3D tracers makes the problem of separating the high-$z$ component more tractable.

% -------------------------- S5: Conclusions -------------------------- %

\section{Conclusions} \label{sec:conclusions}

In this work, we develop the modelling framework for the NIRB signals sourced by Pop~II and Pop~III star-forming galaxies at $z>5$. We leverage a semi-empirical approach to build our model on top of physically-motivated prescriptions of galaxy evolution and star formation under feedback regulation, and calibrate them to observations of high-$z$ galaxies. Using our model, we analyse how the formation histories of first stars may be revealed by measuring the spatial and spectral properties of the NIRB. 

Our main findings can be summarized as follows:
\begin{enumerate}
    \item Using a collection of variations in Pop~II and Pop~III SFHs derived from our model, we reinforce the modelling of the contribution to the NIRB from high-$z$ star-forming galaxies by characterizing the dependence of its shape and amplitude on physics of star formation and galaxy evolution. We find little difference in the predicted contribution of Pop~II stars to the NIRB, given the uncertainty in the SFE allowed by constraints on the faint-end slope of galaxy UVLFs. The Pop~III SFH, on the contrary, is highly uncertain and sensitive to the LW feedback from both Pop~II and Pop~III stars themselves, leading to substantial variations in their imprints on the NIRB. 
    
    \item Depending on exact SFHs and detailed properties of Pop~III stars such as the IMF, they are expected to leave characteristic spectral signatures on the NIRB at wavelengths redward of $1\,\mu$m due to their strong Ly$\alpha$ emission. In our optimistic models with efficient formation of massive Pop~III stars, such signatures can be as strong as up to a few tens of percent of the fluctuations sourced by Pop~II stars, making the NIRB a promising probe of the first stars. Spatial information of the NIRB, such as the shape of the power spectrum, can also shed light on the physics of the first stars, including effects of various feedback processes and the escape of LyC photons. 
    
    \item Forthcoming space missions like SPHEREx and CDIM can quantify the NIRB fluctuations contributed by high-$z$ galaxies, and thereby placing interesting constraints on the Pop~III SFH that is difficult to measure by observing individual galaxies. Even though only optimistic models where massive Pop~III stars of $\gtrsim 100\,M_\odot$ form at a high efficiency of order 0.1--1 in minihaloes (resulting in a peak Pop~III SFRD as high as $\sim 10^{-3}\,M_\odot\,\mathrm{yr^{-1}\,Mpc^{-3}}$, or $\dot{M}^{\rm III}_* \sim 10^{-3}\,M_\odot\,\mathrm{yr}^{-1}$ in individual minihaloes) may be probed in the SPHEREx deep field, ruling out or disfavouring such extremely scenarios with SPHEREx would still be extremely interesting. With better surface brightness sensitivity, the CDIM medium-field survey has the better chance to inspect a larger subset of plausible Pop~III models with less efficient star formation and/or less top-heavy IMFs. 
    
    \item Any constraints on the first stars from NIRB measurements can have interesting implications for other EoR observables, including the global reionization history, 21-cm signal, and the CMB. In the future, joint analyses of all these probes will provide the best opportunity for overcoming observational systematics such as foreground contamination and studying the first stars from an angle different from, and complementary to, the traditional approach of observing individual galaxies. 
\end{enumerate}

\section*{Acknowledgments}
The authors would like to thank Lluis Mas-Ribas for providing updated models of extended Ly$\alpha$ emission and comments on the early draft, as well as Jamie Bock, Tzu-Ching Chang, Asantha Cooray, Olivier Dor\'{e}, Chang Feng, Caroline Heneka, and Adam Lidz for helpful discussion about SPHEREx and CDIM instruments and scientific implications. G.S. is indebted to David and Barbara Groce for the provision of travel funds. J.M. acknowledges support from a CITA National fellowship. S.R.F. was supported by the National Science Foundation through award AST-1812458. In addition, S.R.F. was directly supported by the NASA Solar System Exploration Research Virtual Institute cooperative agreement number 80ARC017M0006. S.R.F. also acknowledges a NASA contract supporting the ``WFIRST Extragalactic Potential Observations (EXPO) Science Investigation Team" (15-WFIRST15-0004), administered by GSFC. 

\section*{Data Availability}
No new data were generated or analysed in support of this research.



\bibliographystyle{mnras}
\bibliography{nirb} % if your bibtex file is called example.bib

%\begin{thebibliography}{99}
%\end{thebibliography}

%%%%%%%%%%%%%%%%%%%%%%%%%%%%%%%%%%%%%%%%%%%%%%%%%%

%%%%%%%%%%%%%%%%% APPENDICES %%%%%%%%%%%%%%%%%%%%%

%%%%%%%%%%%%%%%%%%%%%%%%%%%%%%%%%%%%%%%%%%%%%%%%%%

% Don't change these lines
\bsp	% typesetting comment
\label{lastpage}
\end{document}